\documentclass[
 prl,
rsi,%
 amsmath,amssymb,
 reprint,%
superscriptaddress
]{revtex4-2}
\usepackage{graphicx}% Include figure files
\usepackage{dcolumn}% Align table columns on decimal point
\usepackage{bm}% bold math
\usepackage{color}
\usepackage{xcolor}
\usepackage{lineno}
\usepackage{amssymb}
\usepackage{amsmath}
\usepackage{mathtools}
\usepackage{ulem}
\begin{document}
\preprint{APS/123-QED}
\title{Miniature work-to-work converter engine powered by motor protein}
%Feedback controlled micro-engine powered by motor protein\\
%\textcolor{blue}{kinesin-powered micro-engine yields substantial/significant performance boost\\
%Significant boost in micro-engine performance obtained from motor-optical trap setup}\\
%Performance boost in colloidal microengine powered by motor protein / 
%Significant boost in performance of colloidal microengine powered by motor protein}% Force line breaks with \\}
%\title{A microscale engine powered by motor}% Force line breaks with \\

%\title{Work-to-work converter engine powered by motor}
\author{Suraj Deshmukh}
\affiliation{Department of Physics, Savitribai Phule Pune University, Pune 411007, India}
\affiliation{Department of Physics, Indian Institute of Science Education and Research, Bhopal 462066, India}
\author{Sougata Guha}
\affiliation{Dipartimento di Fisica, Università degli Studi di Napoli Federico II, and INFN Napoli, Complesso Universitario di Monte Sant'Angelo, 80126 Naples, Italy}
%\affiliation{INFN Napoli, Complesso Universitario di Monte S. Angelo, Napoli 80126, Italy}
\author{Basudha Roy}
\affiliation{Department of Physics, Savitribai Phule Pune University, Pune 411007, India}
\author{Shivprasad Patil}
\affiliation{Department of Physics, Indian Institute of Science Education and Research, Pune 411008, India}

\author{Arnab Saha}
\affiliation{Department of Physics, University of Calcutta, Kolkata 700009, India}
\author{Sudipto Muhuri}
\email{sudiptomuhuri@uohyd.ac.in}
\affiliation{School of Physics, University of Hyderabad, Hyderabad 500046, India}
\affiliation{Department of Physics, Savitribai Phule Pune University, Pune 411007, India}

            
% Please add a significance statement to explain the relevance of your work

%\significancestatement{
%Pioneering works related to micro-manipulation techniques using optical tweezers prepared the ground for the experimental realization of microscale engines based on the principle of conversion of heat to work. However since the typical work output per cycle is of the order of $k_b T$ or less, their practical utility is hindered significantly. The novel feedback controlled microscale engine that we envisage is a work-to-work converter which is capable of harnessing the activity of a motor protein into work output. We illustrate how a motor state dependent feedback protocol can lead to engine performance which vastly supersedes the performance of microscale engines experimentally realized so far.}

%\equalauthors{\textsuperscript{$\dagger$}These authors contributed equally}
%\correspondingauthor{\textsuperscript{*}To whom correspondence should be addressed. E-mail: sudiptomuhuri@uoh.ac.in}

% At least three keywords are required at submission. Please provide three to five keywords, separated by the pipe symbol.
%\keywords{Micro heat engines | Molecular motors | Stochastic thermodynamics $|$ First passage times } 

\begin{abstract}

Designing a miniature microscale engine that can override the role of thermal fluctuations has remained elusive and is an important open challenge. 
%Hitherto, conventional colloidal microscale engine have relied upon the working principle of conversion of heat from (a)thermal bath into work output of the engine, resulting in work output which is of the order of thermal fluctuations ($k_bT$).
%For conventional colloidal microscopic engines, the underlying working principle is based upon extraction of heat from a thermal bath and converting it into work output of the engine. The performance of such microscale engines results in a work output per cycle which is of the order of thermal fluctuations ($k_bT$). 
%The fact that the work output per cycle is $\sim k_bT $, hinders the possibility of fabricating microdevice which can have practical usage. 
Here we provide the design and theoretical framework for a unique information-based engine -- a work-to-work converter -- comprising a sub-micron size bead and motor protein-microtubule (MT) complex in an optical trap setup. %
%This microscale engine is capable of harnessing and converting the movement of a single motor protein into work output in cyclic fashion.
We demonstrate how by implementing a simple motor protein state-dependent feedback protocol of the optical trap stiffness, this engine is able to harness and convert the movement of a motor protein into work output. Unlike other conventional microengines, the fidelity and performance of this engine is determined by the stochasticity of motor (un)binding characteristics. We obtain an analytical form of the work distribution function, average work output and average power output, providing quantitative predictions for engine performance which are validated by stochastic simulations. 
Remarkably, the average work output per cycle is at least an order of magnitude higher than the thermal fluctuations and supersedes the performance of other microscale engines realized so far. 
%The fidelity and engine characteristics of the proposed engine are controlled by the stochasticity of motor (un)binding characteristics to MT and not by the underlying bath properties. 
%This engine can be experimentally realized by using a system comprising a colloidal bead and MT-motor complex in an optical trap setup. %The performance of this miniature machine can be suitably tuned and optimized by varying motor velocity and optical trap stiffness. \textcolor{blue}{SG: Must reduce the abstract size heavily}%This microengine is a promising potential prototype for fabricating high performance bio-compatible microdevice in future.

\end{abstract}

\maketitle

%\section{Introduction}

Molecular motors, the naturally occurring miniature machines, are ubiquitous in eukaryotic cells \cite{howard2002mechanics}. These tiny entities utilize stored chemical energy in the form of Adenosine triphosphate (ATP)  to generate directed motion and perform work, accomplishing an amazing variety of complex tasks within the confines of the cell \cite{phillips2012physical, howard2002mechanics, oster2002brownian}. The underlying working mechanism at play for molecular motors is the principle of {\it Brownian ratchets} which involves rectifying and trapping favourable fluctuations for inducing mechanical force and movement \cite{julicher1997modeling,reimann2002brownian,ait2003brownian}.
 Inspired by these naturally occurring miniature machines, the quest for designing and realizing powerful and robust artificial machines which are capable of performing work at microscales has remained a conceptual challenge.
 %Conceptualization of such microscopic engines much like their macroscopic counterparts are based on the principle of conversion of heat or chemical energy into mechanical work \cite{ schmiedl2007efficiency,rana2014single,verley2014universal, verley2014unlikely, blickle2012realization,martinez2016brownian,krishnamurthy2016micrometre,roy2021tuning,krishnamurthy2023overcoming}.
 
In a set of landmark experiments, microscale Carnot and Stirling engines operating between two thermal baths have been realized using micron size colloid in optical trap setup \cite{blickle2012realization,martinez2016brownian,ciliberto2017experiments}. However, a fundamental drawback for these microengines is that the work output per cycle is less than $k_BT$ which hinders their practical utility. Efforts to enhance engine performance have taken recourse to different strategies. One strategy has focused on improvement  by engineering the bath properties \cite{krishnamurthy2016micrometre,krishnamurthy2023overcoming, roy2021tuning,saha2019stochastic}. It has been illustrated that more thermodynamic work can be extracted if the thermal reservoirs  are replaced by bacterial baths where live motile bacteria collide incessantly with the system particle, producing active, non-equilibrium fluctuations \cite{krishnamurthy2016micrometre,krishnamurthy2023overcoming,saha2019stochastic, roy2021tuning}. It is intuitively expected that the performance of microengines can be further enhanced if the information regarding the {\it state of the system} is known {\it a priori} \cite{szilard1929entropieverminderung,parrondo2015thermodynamics,cao2009thermodynamics,tohme2024gambling, du2024performance,saha2023information, malgaretti2022szilard, paneru2018lossless, ribezzi2019large, paneru2022colossal}. The idea of {\it Maxwell's demon} \cite{leff2002maxwell} has been rekindled to design information engines which are able to utilize the information pertaining to the state of the system to extract work from thermal fluctuations \cite{paneru2018lossless,paneru2022colossal,ribezzi2019large}. Such a strategy has been ingeniously adopted in a DNA hairpin experiment, wherein by making use of the knowledge of state of the DNA (folded or unfolded), it has been demonstrated that mechanical work can  be extracted in cyclic fashion \cite{ribezzi2019large}.  %to extract work by  demon has been realized in a DNA hairpin pulling experiment, wherein it has been demonstarted that 

\begin{figure}[t]
    \centering
    \includegraphics[width=\linewidth]{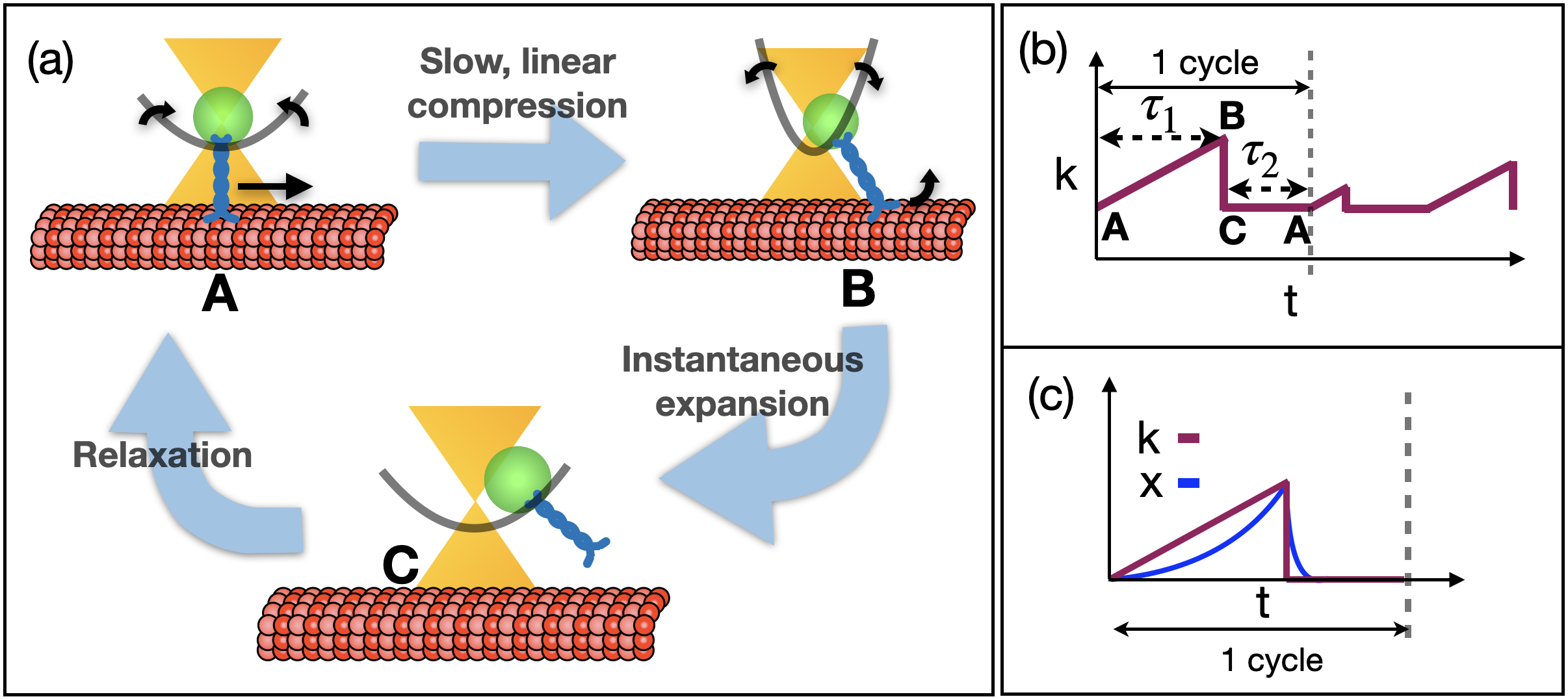}%{Schematic_engine_main_22.png}%{Schematic_engine_main.png}
    \caption{Engine Cycle: (a) At \textit{A}, motor attaches at optical trap center. For step \textit{AB}, trap stiffness $k_t$ varies linearly such that $k_t(t) = k_o + \mu t$. At \textit{B}, motor detaches from MT and $k_t$ is instantaneously reduced to $k_o$, corresponding to the step \textit{BC}. For step \textit{CA}, the detached motor relaxes to the center of optical trap and stays there until next motor attachment event occurs. This sequence of events completes one engine cycle.
    %The bead waits here ($\bar{x}=0$) for a time $\tau_m$ until the next binding event is realized reaching \textit{instance A}, thereby completing one cycle. 
    (b) Trap stiffness variation with time : $k_t$ vs $t$ (c) Variation of bead position $x$ with $t$, and its correspondence with $k_t$ vs $t$ for one complete engine cycle.}
    %. position $x$ vs $t$ and its comparison corresponding $x$The corresponding One complete cycle of protocol, along with mean position of bead.}
    %Schematic of a motor attached bead/colloid at being transported on a microtubule (MT) filament in an optical trap. $\tau$ is the runtime of the motor until detachment from MT. $\tau_m$ is the time interval after which motor attaches to MT. The total cycle time $t = \tau + \tau_m$.
    
    \label{fig1}
\end{figure}

Notably, realization of all such microengines, much like their macroscopic counterparts, are based on the principle of conversion of heat energy from an underlying (a)thermal bath into mechanical work \cite{ schmiedl2007efficiency,rana2014single,verley2014universal, verley2014unlikely, blickle2012realization,martinez2016brownian,krishnamurthy2016micrometre,roy2021tuning,krishnamurthy2023overcoming, prlroy}.
%\cite{blickle2012realization,martinez2016brownian,ciliberto2017experiments, krishnamurthy2016micrometre,krishnamurthy2023overcoming,saha2019stochastic, ritort}. 
In contrast to such engines, we provide the template and design for a {\it feedback controlled} work-to-work converter engine which is powered by a motor protein. Conceptually, the functionality of this engine is achieved by converting the work done by a motor protein into work output of the engine, in a feedback controlled manner. 
While for other conventional microengines, the work output and engine characteristics is strongly influenced by (a)thermal fluctuations of the bath, we show that for this work-to-work converter, the engine performance and its fidelity is primarily determined by the stochasticity of the motor (un)binding processes and {\it not the fluctuations of the underlying (a)thermal bath}. 
\emph{Model} --- The engine that we envisage comprises of a sub-micron size bead-{\it kinesin} motor complex, in a thermal bath. The bead is subjected to a time-varying, feedback-controlled optical trap potential and a driving force due to the action of the motor protein which stochastically binds, walks, and unbinds to an underlying microtubule (MT) filament. 
%For this system whenever the motor protein binds to the underlying MT filament, the bead gets dragged along the MT, experiencing a pulling force due to the motor and an opposing restoring force due to the harmonic potential of the optical trap. 
%Thus effectively, the motor performs work on the system. 
The engine cycle comprises of stochastic attachment of motor to MT, its motion along MT while it drags along the bead, stochastic detachment of the motor from MT and subsequent relaxation of the bead to the trap center (Fig.\ref{fig1}a). The feedback control operates in such a manner that whenever the motor attaches to the MT, the trap stiffness $k_t$, increases linearly as $k_t(t) = k_o + \mu t$ until the instant when the motor detaches from MT. We choose the linear increase of $k_t$ such that $\mu t \ll k_o$. At the instant of a detachment event of motor, $k_t$ is instantaneously reduced to $k_o$ and it continues to remains so until the next event of motor attachment (Fig.\ref{fig1}b). Fig.\ref{fig1}c depicts the variation of trap stiffness and bead displacement as a function of time in one cycle.

First, we propose and analyze an effective 1D model for the system, wherein the bead movement is considered to be along the axis of the MT filament \cite{sundararajan2024theoretical, kunwar2011mechanical}. Subsequently, we discuss engine performance for a generalized model which takes into account the full 2D movement of the bead in an optical trap setup \cite{ashkin1992biophysj,fisher2005pnas,khataee2019prl,pyrpassopoulos2020biophysj,sundararajan2024theoretical}.

%Let $x(t)$ be the instantaneous position of the bead and $x_m(t)$  the position of the motor on the MT. If the rest length is set to zero then, the particle experiences an instantaneous driving  force, 
%\begin{equation}
%f(t) =  k_m \left[ x_m(t) - x (t) \right] 
%\label{eq-f1}
%\end{equation}
%Owing to Newton's third law, the same force is felt by the motor in the opposite direction. 
For the 1D model, when motor is attached to the MT filament, the corresponding Langevin dynamics for the bead position $x(t)$, in the overdamped limit is,
\begin{equation}
\gamma \dot x = -k_t(t) x +   f(t) + \xi(t),
\label{eq-langevin}
\end{equation}
Here, $f(t)$ is the force due to motor, $\gamma$ is the friction coefficient and $\xi(t)$ is the random force experienced by the bead due to thermal fluctuations of the bath. Kinesin motor can be considered as a harmonic spring with an effective spring constant $k_m$. Therefore, $f(t) =  k_m \left[ x_m(t) - x (t) \right]$, where $x_m(t)$ is the instantaneous position of the motor with respect to the optical trap center \cite{coppin1997load}. 
%For usual thermal bath, the random force satisfies the usual property of a thermal bath in equilibrium, i.e.,  $\overline \xi  = 0$ and $\overline{\xi(t_1)\xi(t_2)} = 2 \gamma k_b T \delta(t_1 - t_2)$, where $\overline{(\dots)}$ denotes thermal average over the bath degrees of freedom. When the motor is not attached to the MT filament, $f(t) =0$ and the dynamics of the bead is simply described by an overdamped Langevin equation for particle in a one-dimensional harmonic potential with a trap stiffness $k_t$. 
%We specify a time-dependent form of the optical trap stiffness, such that, $k_t(t) = k_o + \mu t$,
%\begin{eqnarray}
% $k_t(t) = k_o + \mu t$,
 %\label{eq-kt}
 %\end{eqnarray}
 %whenever the motor is attached to the MT filament, while $k_t = k_o$ whenever the motor is in detached state. Indeed this is the prescribed {\it feedback control} mechanism which results in a net work output by  the engine. 
 The average velocity of the motor on the MT filament when it is subject to a load force $f$ is well described by a linear force-velocity relation \cite{klumpp2005cooperative,muller2008tug,puri2019dynein, svoboda1994force, Visscher1999}. This linear relation can be expressed as \cite{klumpp2005cooperative,svoboda1994force},
 %Assuming a linear force-velocity relation for the motor, the average velocity of the motor on the MT filament when it experiences a force $f$, can be expressed as \textcolor{red}{\cite{klumpp2005cooperative,muller2008tug,muhuri2010lattice,puri2019dynein, svoboda1994force}}, 
 \begin{equation}
 \dot x_m = v_{o}\left( 1 - f/f_s \right),
 \label{eq-vm}
 \end{equation}
 where $v_o$ is the average velocity of motor at zero load force and $f_s$ is the characteristic stall force for the motor.

 % For this system comprising of micron size bead in solution, there are three intrinsic time scale; (i) Thermal relaxation time scale $\tau_c$, the hydrodynamic relaxation time scale of the bead to relax to the trap center $\tau_b$, and the (iii) the timescale of motor (un)binding and movement, $\tau_m$ which are well separated such that $\tau_c << \tau_b <<  \tau_m$.  which have a typical range of  $\tau_m \sim (10^{-1} - 10^{1})$ s. We note that these time scale are well separated for our system such that $\tau_c << \tau_b <<  \tau_m$. Thus for any configuration of the motor in the attached state at any instant, from Eq.\ref{eq-langevin}, it follows that the displacement of the particle from the trap center, averaged over the thermal bath degrees of freedom, $\overline{x}$, satisfies the relation
%  \begin{equation}
%   k_t \overline x = k_m(x_m - \overline x ) = f(t).
%   \label{av-x}
%  \end{equation}
%\begin{equation}
%\dot x_m = v_{o}\left( 1 - f/f_s \right),
%\label{eq-vm}
%\end{equation}
%where $x_m$ is the displacement of the bead from the optical trap center, $v_o$ is the velocity of motor at zero load force and $f_s$ is the characteristic stall force for the motor.
% The Eq.\ref{eq-langevin} together with Eq.\ref{eq-vm} are coupled via the driving force $f(t)$. 
\emph{Results} ---
For this system the three intrinsic time scales, e.g., (i) mean collision time scale of bead with fluid molecule of the bath $\tau_c$, (ii) the hydrodynamic relaxation time scale of the bead to relax to the trap center $\tau_b$, and (iii) the time scale of motor (un)binding and movement, $\tau_m$ are well separated such that $\tau_c \ll \tau_b \ll  \tau_m$ \cite{timescale}. 
% For sub-micron size bead, $\tau_c \sim 10^{-18}s, \tau_b \sim 10^{-4} s ( for $k_o = 0.005 pN Nm^{-1}) and $\tau_m has a typical range of  $\tau_m \sim (10^{-2} - 10^{1}) s $
Consequently, the thermally averaged bead position $\overline x(t)$ satisfies the force balance relation, $k_t \overline x = k_m(x_m - \overline x )$ at all instants of time \cite{sundararajan2024theoretical}.
%This relation is a statement of force balance condition being satisfied at all instant of time.
 %\textcolor{red}{Note that we ignored the possible fluctuations in motor positions as the fluctuation in motor step size ($\sim0.08~nm$) is much smaller than the mean step size ($8~nm$) \cite{fehr2008kinesin}.}
%include in reference the discussion on time scalkes
In order to obtain a functional form of $\overline x(t)$, we first take the thermal average of Eq.\ref{eq-langevin}. Then we take the Laplace transform to obtain $\overline x(s)$. Invoking the limit of {\it fast} hydrodynamic relaxation and {\it slow} variation of $k_t$ and taking inverse Laplace transform, we obtain 
\begin{equation}
   \overline x(t)  = \frac{f_s}{k_o}(1- e^{-\alpha t})
\end{equation}
 Here, $\alpha = \left(\frac{v_o}{f_s}\right)\left(\frac{k_o k_m}{k_o + k_m}\right)$ is an inverse time scale that is determined by trap stiffness and motor property. Detailed calculations are presented in section I of Supplementary Material (SM).

The unbinding kinetics of kinesin motor from the MT filaments has a general form, 
$\epsilon = \epsilon_o e ^{f/f_m}$,
where, $\epsilon_{o}$ is the unbinding rate of a single motor in the absence of load force $f$, while $f_m$ is a characteristic force scale associated with the unbinding  process \cite{brenner2020force,klumpp2005cooperative}. 
In order to calculate the distribution of the thermodynamic quantities as a function of the stochasticity of the (un)binding process, we need to know the  Probability distribution function (PDF) of the runtime of motor $\tau_1$ - the duration for which the motor is attached to MT and the duration of time for which the motor remain unbound $\tau_2$ during a particular engine cycle. 
%Subsequently, we can then obtain the expressions for the thermodynamic quantities, e.g., work done, power, and efficiency , {\it averaged over the stochasticity}.
The PDF for $\tau_2$ has a form $P(\tau_2) = \pi_o e^{-\pi_o \tau_2}$, where $\pi_o$ is the binding rate of a single motor protein to MT \cite{klumpp2005cooperative, leduc2004cooperative}. 
%In order, to obtain the expression for the probability distribution function of runtime of motor, $P(\tau_1)$, we proceed as follows: We define $S(\tau_1)$ as the {\it Survival time probability} distribution of the motor i.e., the probability that the motor remains attached to the filament after time $ t = \tau_1$, starting from the initial position of the optical trap center at time $t=0$. 
%The time evolution of $S(t)$ can then be expressed as, $ \frac{dS}{dt} = -\epsilon(t)S$. Using the relation $P(t)=-\frac{dS}{dt}$, the final form of 
The PDF of the runtime of the motor can be expressed as \cite{sundararajan2024theoretical},
\begin{equation}
    P(\tau_1)=  \epsilon_{o} \exp \left[ \frac{f_s}{f_m} \left( 1 - e ^{- \alpha \tau_1} \right )\right ]  \exp \left [ -\int_{0}^{\tau_1} \epsilon(t) dt\right ]
\label{eq:PDF-t}
\end{equation}
where the unbinding rate, 
%the explicit functional form of the unbinding rate as function of time is 
%\begin{equation}
$\epsilon (t) = \epsilon_{o} \exp \left[ \frac{f_s}{f_m} \left( 1 - e ^{- \alpha t}\right) \right]$. While the average runtime $\langle \tau_1 \rangle$ ($\langle ... \rangle$ denotes average over different runtimes) is always a monotonically decreasing function of $\alpha$, $P(\tau_1)$ changes its behaviour from a monotonically decreasing function of $\tau_1$ to exhibiting a peak at $\tau_1=t_o$, beyond a value of  $\alpha = \alpha_c$ (Fig.S1). 
%The value $t_o$  can be obtained by setting ${(dP/d \tau_1)}_{\tau_1= t_o} = 0$. 
The resultant expression of $t_o$ can be expressed in terms of Lambert-W function as (see section II of SM and Fig.S2), 
%\textcolor{red}{SD: The result is expressible in a closed form by utilizing the Lamberts-W function $W(x)$ which is inverse function of $x = We^W$.
\begin{equation}
t_o(\alpha) = \frac{1}{\alpha}\left[\ln\left(\frac{\alpha f_s}{\epsilon_o f_me^\frac{f_s}{f_m}}\right) + W_o\left(\frac{e^{\frac{f_s}{f_m}}\epsilon_o}{\alpha}\right)\right]
%\label{maxPtau} 
\end{equation}

%\section{Results}
%\subsection{Expressions of thermodynamic quantities}
%In order to obtain the explicit expressions for the average thermodynamic quantities for the engine, we proceed as follows. First we  identify that corresponding to the restoring force due to optical trap spring, we can associate a potential energy of the form $U(x) = \frac{1}{2} k_t(t) x^2$. Indeed the restoring force due to this potential energy corresponds to a conservative force in contrast to the driving force due to the motor.  Then the Langevin Equation describing the system  can be recast in the form of the first law of thermodynamics \cite{peliti2021stochastic}. In order to see the connection, we integrate  Eq.(\ref{eq-langevin}) for all possible value of the position of the Brownian particle, $x$ corresponding to a particular path. The corresponding form of the integral expression then can be cast in the following form:
{\it Engine Cycle} --- 
We obtain the expression for the work output of the system by recasting the Langevin equation (Eq.\ref{eq-langevin}) in the form of the first law of thermodynamics \cite{peliti2021stochastic}.
%\textcolor{blue}{The potential energy of the bead is $U(x) = \frac{1}{2} k_t(t) x^2$. The corresponding Langevin Equation can be recast in the form of the first law of thermodynamics , allowing us to obtain the expression for the work output of the system \cite{peliti2021stochastic}.}
%\textcolor{blue}{SD:can we avoid "typical"?}
The engine cycle begins with the motor attachment event at $\overline{x} = 0$. The potential energy of the bead is given by $U = \frac{1}{2} k_t(t) \overline{x^2}\simeq\frac{1}{2} k_t(t) \overline{x}^2$. This assumption is reasonable as long as the average work output of the system in one cycle is much larger than the contribution due to thermal fluctuations (see section III of SM). Indeed, our results for the work output (discussed later) confirm the veracity of the assumption that we have invoked. For the step {\it AB}, work done on the system $\Delta W^{(AB)}_{c}  = \frac{\mu}{2} \int_{o}^{\tau_1} \overline{x}^{2} dt$ while for step BC, work done by the system is $\Delta W^{(BC)}_c= -\frac{1}{2}\mu \tau_1 \overline{x}^{2}(\tau_1)$.
%(\textcolor{red}{SG: Add the calculation in SI}). When the motor detaches at $t = \tau_1$, the trap stiffness is instantaneously changed from $k_o + \mu \tau_1$ to $k_o$. The corresponding expression for work output for this step (path BC) is \textcolor{red}{$\Delta W^{(BC)}_c= -\frac{1}{2}\mu \tau_1 {\overline x}^{2}(\tau_1)$}. 
%\textcolor{blue}{In order to obtain the explicit expression for the work output, we assume that $\Delta W^{(AB)}_{c}  \simeq \frac{\mu}{2} \int_{o}^{\tau_1} { \overline x }^{2} dt$ and $\Delta W^{(BC)}_c \simeq -\frac{1}{2}\mu \tau_1 {\overline x}^{2}(\tau_1)$}.
For the step CA, $\Delta W^{(CA)}_{c}= 0$.
The corresponding expression for net work output in an engine cycle for a motor runtime $\tau_1$ is (see section IV(A) of SM for details),
\begin{equation}
W_c = \left(\frac{-\mu f_s^3}{v_o k_o^3}\right)\left[ \frac{3}{4} + \frac{1}{4}(2\alpha \tau_1 +1 )e^{-2\alpha \tau_1} - (\alpha \tau_1 +1)e^{-\alpha \tau_1}\right]
\label{W-total}
\end{equation}

%The maximum possible displacement of the motor and consequently of the bead from the trap center is determined by the stall force $f_s$(\textcolor{red}{SG:Do we need this sentence here??}). 
%\end{equation}
%Since work output of the engine is function of the run time $\tau_1$, therefore, 
The PDF of the work output can be expressed as
\begin{equation}
P(W_c) = J(\tau_1)P(\tau_1)=\left|\frac{d \tau_1}{d W_c}\right| P(\tau_1)
\label{Pw}
\end{equation}
 Here, the explicit expression for the Jacobian is, $ J(\tau_1) = \left(\frac{k_o}{f_s}\right)^2\left[\frac{e^{\alpha\tau_1}}{\alpha\mu\tau_1(1-e^{-\alpha\tau_1})}\right]$. 
The average work done in a cycle over different realization of runtime $\tau_1$ is simply, 
\begin{equation}
\langle W_c \rangle = \int_0^{\infty} W_c~ P(\tau_1) d\tau_1
\label{eq:Wc1}
\end{equation}
The corresponding expression for average work done by the motor on the system per cycle can be expressed as 
$\langle W_m \rangle = - \frac{1}{2} k_o v_{o}^{2} \left[\frac{I_{m}(\alpha)}{\alpha^{2}}\right]$, where 
$I_{m}(\alpha) = \left \langle 1 + e^{-2\alpha\tau_1} - 2 e^{-\alpha\tau_1} \right \rangle$ (see section IV(A) and Fig.S3 of SM).

In the limit of $\tau_1 \rightarrow \infty$, which corresponds to maximum value of work output $W_{max} = \left(-\frac{3\mu f_s^3}{4v_o k_o^3}\right)$,  
 %The expression for the maximum work output, $W_{max}$, obtained by setting $\tau_1 \rightarrow \infty$, reads as 
%\begin{equation}
%$W_{max} = \left(-\frac{3\mu f_s^3}{4v_o k_o^3}\right)$.
 $J(\tau_1)P(\tau_1) \sim \exp \left(\alpha\tau_1 - \epsilon_o \tau_1 e^{\frac{f_s}{f_m}} \right)$. This implies that beyond a threshold value of $\alpha_c = \epsilon_o e^{\frac{f_s}{f_m}}$, $P(W_c)$ would exhibit non-monotonic behaviour and it would also exhibit divergence at $W_c = W_{max}$. 

In the limit of $\alpha \langle \tau_1 \rangle \gg 1$, $\langle W_c \rangle \simeq W_{max}$.
In contrast, for $\alpha \langle \tau_1 \rangle \ll 1$,  $P(W_c)$ is a monotonically decreasing function and assumes the shape of Weibull distribution function of the form (see section IV(B) of SM),
\begin{equation}
P(W_c) = \frac{\epsilon_o}{3}{\left(\frac{3}{\mu v_o^2 W_c^2}\right)}^{1/3}\exp\left[ -\epsilon_o {\left(\frac{3\left|W_c\right|}{\mu v_o^2}\right)}^{1/3}\right]
\end{equation}
%In this limit, $P(W_c)$ is a monotonically decreasing function of $W_c$. 
In this limit, the corresponding expression for average work output (Fig.S4) in a cycle is, $\langle W_c \rangle = - 2\left( \frac{\mu v_o^2}{\epsilon_o^{3}}\right)$.
The average power output per cycle $P_o$, is defined as ratio of average work output to average cycle time. In the limit of $\alpha \langle \tau_1 \rangle \ll 1$, it can be expressed as,
$P_o \simeq  -\frac{2\mu v_{o}^{2}\pi_o }{\epsilon_o^{2}(\epsilon_o + \pi_o)}$.

\begin{figure}[t]
    \centering
\includegraphics[width=\linewidth]{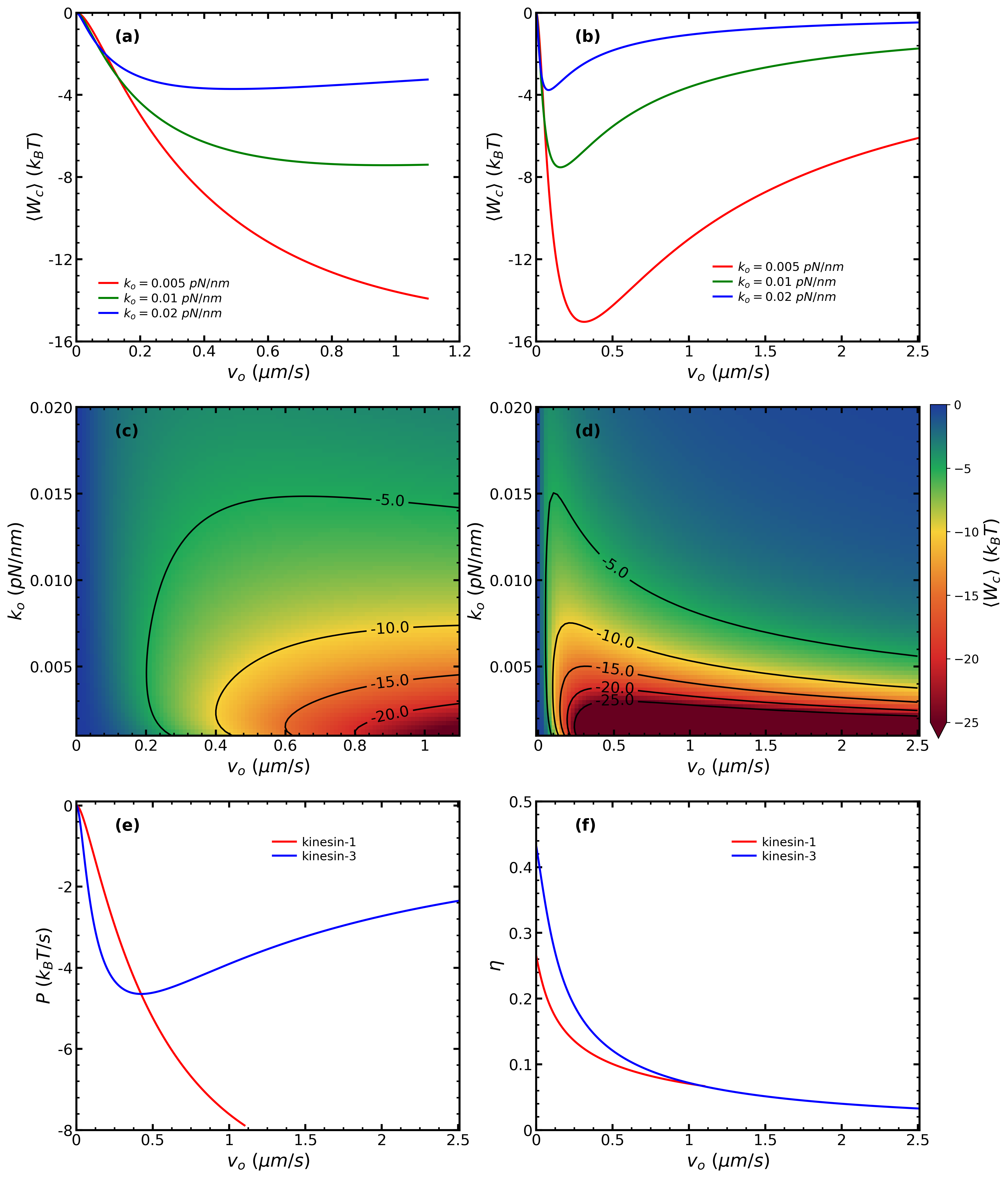}%{Fig2-tight.pdf}
    \caption{Performance of  {\it kinesin-1} and {\it kinesin-3} driven engine: $\langle W_c\rangle$ vs $v_o$ for (a) {\it kinesin-1} and (b) {\it kinesin-3}. Contour plot of $\langle W_c\rangle$ in $(v_o- k_o)$ plane for (c) {\it kinesin-1} and (d) {\it kinesin-3}. (e) Efficiency $\eta$ vs $v_o$ (f) Average power output: $P$ vs $v_o$. For \textit{kinesin-1}, $\epsilon_o = 0.72 ~ s^{-1}$ \cite{soppina2022kinesin}, $f_s = 5.7$ pN \cite{brenner2020force}, and $f_m = 4$ pN \cite{guo2019force}, and for {\it kinesin-3}, $\epsilon_o = 0.23~ s^{-1}$ \cite{soppina2022kinesin}, $f_s = 3$ pN \cite{budaitis2021pathogenic}, and $f_m = 2.7$ pN \cite{budaitis2021pathogenic}. The trap stiffness $k_o = 0.005 ~ pN ~nm^{-1}$ \cite{rai2013molecular,brenner2020force} in (e) and (f).
    %Performance of  {\it kinesin-1} and {\it kinesin-3} driven engine: (a) $\langle W_c\rangle$ vs $v_o$ and (b) Contour plot of $\langle W_c\rangle$ in $(v_o- k_o)$ plane for {\it kinesin-1}. For (a) and (b), $\epsilon_o = 0.72 ~ s^{-1}$, $f_s = 5.7$ pN, and $f_m = 4$ pN \cite{brenner2020force}. (c) $\langle W_c\rangle$ vs $v_o$ and (d) Contour plot of $\langle W_c\rangle$ in $(v_o- k_o)$ plane for {\it kinesin-3}. For (c) and (d), $\epsilon_o = 0.23~ s^{-1}$, $f_s = 3$ pN, and $f_m = 2.7$ pN. \cite{budaitis2021pathogenic,soppina2022kinesin}(e) Efficiency $\eta$ vs $v_o$ (f) Average power output: $P$ vs $v_o$. For (e) and (f), $k_o = 0.005 ~ pN ~nm^{-1}$ .
    }
    \label{fig3}
\end{figure}

%\section{Engine powered by kinesin motor}
\noindent
%One of the primary biological role of kinesin motor proteins in context of cellular processes is intracellular trafficking. kinesin motors stochastically (un)bind to MT filaments and transport cellular cargo from one location to another within the cell. These motor proteins perform their movement on MT filament  by hydrolyzing ATP and converting this chemical energy into mechanical movement on MT. 
%kinesin-1 family of motors are well characterized and studied extensively for their motility and force generation characteristics \cite{brenner2020force,svoboda1994force}. kinesin-1 motors are capable of moving with moderate speeds of  $\sim 1 \mu m s^{-1}$ at saturating ATP concentration under load free conditions. They can sustain relatively high forces, with typical detachment force $f_m \sim 4.7$pN. In contrast, kinesin-3 motors are super-processive, attaining speed of $\sim 2.4 \mu m s^{-1}$ under load free conditions but their ability to sustain forces is relatively poor with $f_m \sim  2.7$ pN \cite{soppina2022kinesin,budaitis2021pathogenic}. So they more readily detach from MT under load force as compared to kinesin-1 motors. 

\emph{Engine powered by kinesin motor} -  Next we study the performance of the engines driven by {\it kinesin-1} and {\it kinesin-3} motors, using the relevant motor parameters known from experiments(see Table I of SM for parameter values).
%For kinesin motors, the velocity $v_o$  can be  varied by  changing the concentration of ATP \cite{howard1989movement}.  
%The optical trap stiffness $k_o$ can be varied by changing the power of the laser. The typical working range of $k_0$  $5 \times 10^{-3} - 10^{-1}$) $pN nm^{-1}$. 
%\textcolor{red}{The list of all the relevant motor parameters measured in experiments is listed in Table-\ref{table1}}. \textcolor{blue}{SD: Table-I goes to supplementry}. 
Fig.\ref{fig3}a and Fig.\ref{fig3}b displays the variation of $\langle W_c\rangle$ as a function of $v_o$, obtained from Eq.\ref{eq:Wc1}, for different values of trap stiffness for \textit{kinesin-1} and \textit{kinesin-3} motors, respectively. In general, for weaker trap stiffness, the work output is higher. Comparison with 1D stochastic simulation which takes into account the stochasticity of the finite size stepping of motor on MT, shows an excellent match with Eq.\ref{eq:Wc1} (Fig.S5). This result points to the insensitivity of $\langle W_c \rangle$ on the stochasticity associated with motor stepping event and motor step size. Remarkably, both for a microengine powered by \textit{kinesin-1} and \textit{kinesin-3} motor, $\langle W_c \rangle $ can easily exceed $12 ~k_BT$ (Fig.\ref{fig3}a and Fig.\ref{fig3}b). 
%\textcolor{blue} {The fact that the typical work output is substantially larger than thermal fluctuations validates the assumption of neglecting the thermal fluctuations of the position of the bead in estimating the work output per cycle.} 
%For this engine, the average power output can be as high as $7 ~k_b T ~s^{-1}$ for \textit{kinesin-1 }when $v_o = 0.8 ~\mu m~ s^{-1}$ (Fig.\ref{fig3}e). 
 %For engine powered by kinesin-1 motor, for a trap stiffness $k_o = 0.005 ~ pN  nm^{-1}$, the average work output is $\sim 12 k_b T$, when $v_o = 0.8 ~ \mu m ~ s^{-1}$ ( Fig.\ref{fig3}a). The corresponding average power output $P \simeq 7 ~ k_b T ~s^{-1}$ ( see Fig.\ref{fig3}e).  For engine powered by kinesin-3 motor, for the same trap stiffness, the average work output for a single is $\sim 15 k_b T$, when $v_o \simeq 0.4 ~\mu m ~ s^{-1}$ ( Fig.\ref{fig3}c). For this case, the corresponding average power output $P \simeq 5 k_b T s^{-1}$. 
Fig.\ref{fig3}c and  Fig.\ref{fig3}d displays contour plot for the work output of the engine powered by \textit{kinesin-1} and \textit{kinesin-3}, respectively, highlighting the non-monotonic behaviour of $\langle W_c \rangle$ as a function of $v_o$ and $k_o$. For this engine, the average power output $P_o$ can be as high as $7 ~k_B T ~s^{-1}$ for \textit{kinesin-1 }when $v_o = 0.8 ~\mu m~ s^{-1}$ (Fig.\ref{fig3}e). Fig.\ref{fig3}f displays the monotonically decreasing nature of the bare efficiency $\eta$, defined as $\eta = \langle W_c\rangle /  \langle W_m\rangle$, as a function of $v_o$. While $\eta$ only captures the efficiency viz-a-viz the work input by the motor, in order to estimate the overall thermodynamic efficiency $\eta_t$, the chemical energy expended by kinesin motors during their ATP hydrolysis has to be accounted for. Since the runlength of the motor in optical trap is $\sim 300-500 ~nm$, and the energy consumed by motor per step of $8 ~nm$ (corresponding to 1 ATP cycle) is $\sim 12~ k_BT$ \cite{rosing1972value, schnitzer1997nature}, the energy consumption per engine cycle is $\sim 450-750 ~ k_B T$, resulting in a typical value of $\eta_t \sim 0.02-0.04$. 

 \begin{figure}[t!]
    \centering
\includegraphics[width=\linewidth]{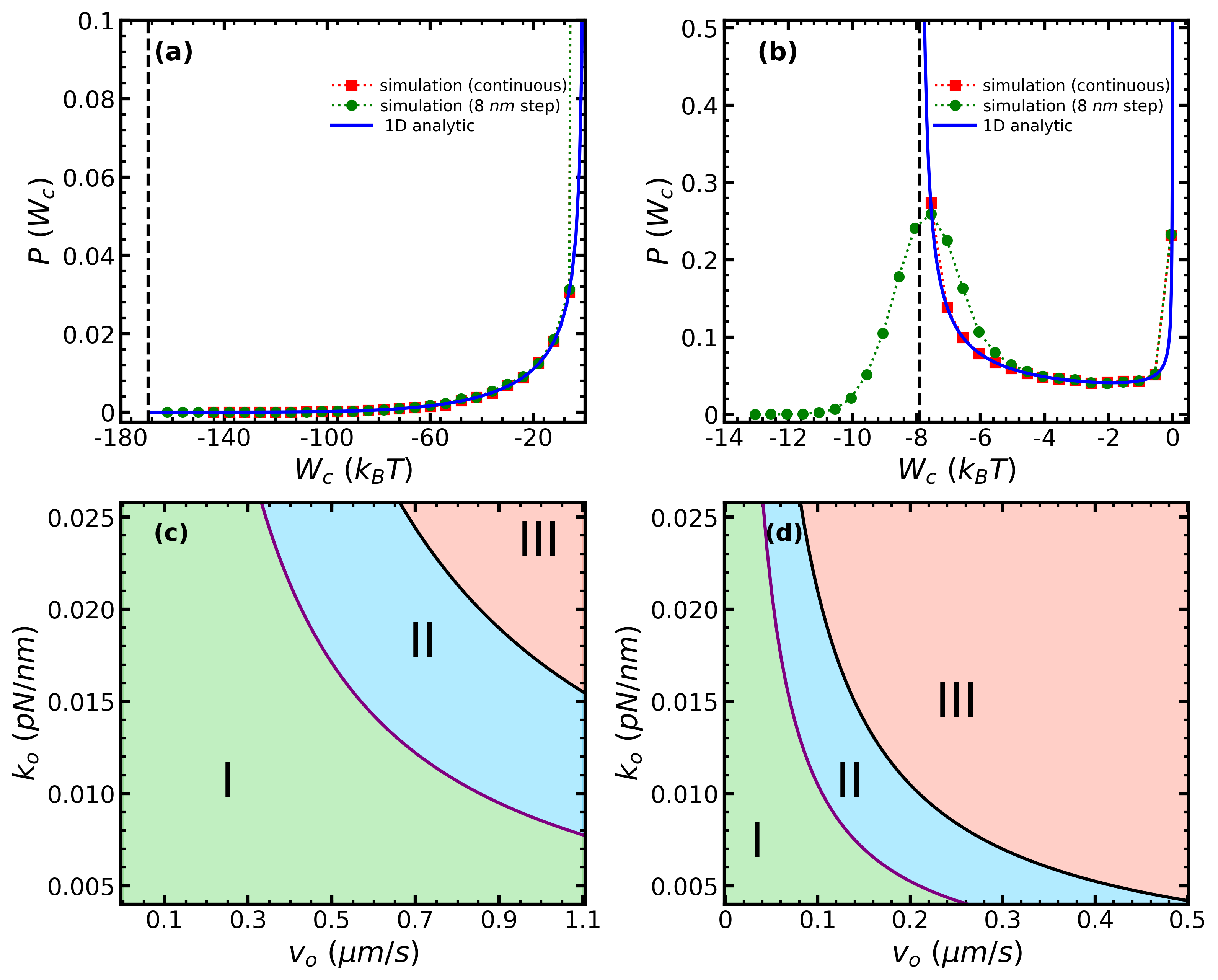}%{Fig3_phase.png}%{Fig3_type2.png}%{work_dist_figure3.png}
    \setlength{\belowcaptionskip}{-8pt}
    \caption{PDF of $W_c$ and its characteristics: (a) and (b) are plots of  P($W_c$) vs $W_c$ for an  engine driven by {\it kinesin-1} and {\it kinesin-3} motor respectively, when $k_o = 0.005~ pN ~nm^{-1}$. For (a), $v_o = 0.8 ~\mu m~s^{-1}$, and for (b),  $v_o = 2.5 ~\mu m~s^{-1}$. Rest of the parameters are same as in Fig.\ref{fig3}. Analytical result of PDF in Eq.\ref{Pw} (blue curve) is compared with 1D stochastic simulations (performed with $10^6$ samples) with discrete motor step size of $8~ nm$ (green circles) and continuous step size (red squares). The maximum work output $W_{max} = - 169.42 ~ k_BT$ for (a), and $W_{max} = - 7.90 ~ k_BT$ for (b).   (c) and (d) corresponds to plot of the characteristics of P($W_c$) for an  engine driven by {\it kinesin-1} and {\it kinesin-3} motor respectively. In Region $(I)$ (green), $P(W_c)$ monotonically decreases to zero. In Region $(II)$ (blue) $P'(W_c) \rightarrow -\infty$ as $W_c \rightarrow W_{max}$. In Region $(III)$ (red) both $P'(W_c) \rightarrow \infty$ and  $P(W_c) \rightarrow \infty$ as $W_c \rightarrow W_{max}$. The purple curve corresponds to $\alpha = \alpha_c/2$ while the black curve corresponds to $\alpha = \alpha_c$.}
    \label{fig-work-dist-1d}
\end{figure}
 %Further, even this definition does not account for the energy cost associated with implementing the feedback process.
 %Using the average displacement corresponding to average work done per cycle, the amount of chemical energy per cycle consumed provides an estimate of $\sim 400-500~ k_b T$. Therefore the maximum thermodynamic efficiency can estimated to to $\eta_t \sim 0.01-0.05$. Further, even this definition does not account for the energy cost associated with implementing the feedback process.

Fig.\ref{fig-work-dist-1d}a and Fig.\ref{fig-work-dist-1d}b displays the PDF of work for \textit{kinesin-1} and \textit{kinesin-3} motor-driven engine, respectively. The PDF exhibits both monotonic and non-monotonic behaviour, depending on the parameter regime (Fig.\ref{fig-work-dist-1d}c and Fig.\ref{fig-work-dist-1d}d). While the comparison of the analytical result in Eq.\ref{Pw} with 1D stochastic simulation is excellent in the monotonic regime, for the non-monotonic regime, the deviations that are observed at higher values of $W_c$ can essentially be attributed to the effect of finite step size of motors (see section V and Fig.S6 of SM). 

\begin{figure}[hbt!]
    \centering
\includegraphics[width=\linewidth]{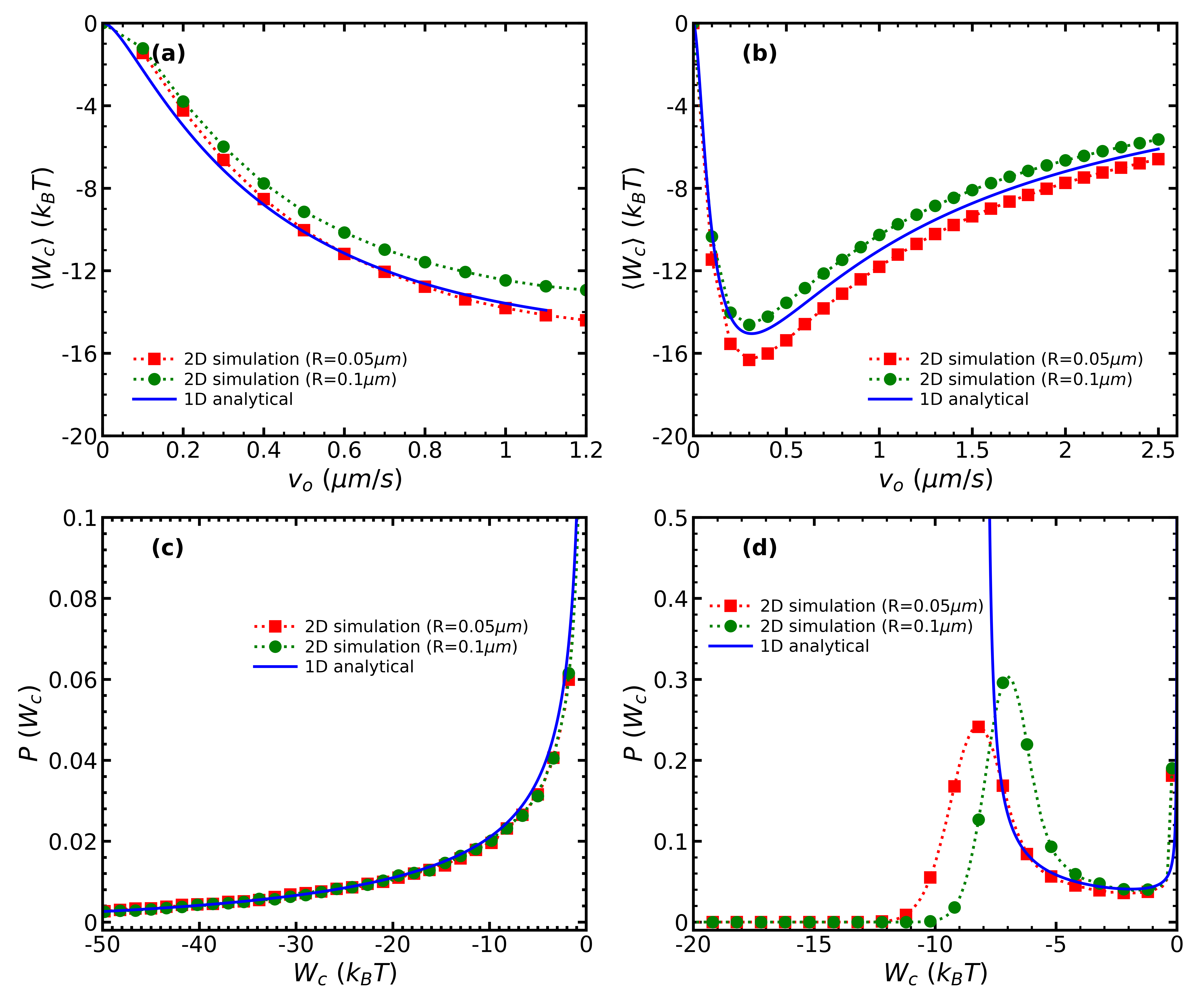}
    \caption{Comparison with 2D stochastic simulations: (a) and (b) correspond to plots of $\langle W_c\rangle$ vs $v_o$ for engine driven by \emph{kinesin-1} and \emph{kinesin-3} motor  respectively. (c)  $P(W_c)$ vs $W_c$ for engine driven by \emph{kinesin-1} motor when $v_o=0.8~\mu m ~s^{-1}$ \cite{howard1989movement}. (d) $P(W_c)$ vs $W_c$ for engine driven by \emph{kinesin-3} when $v_o=2.5~\mu m~s^{-1}$ \cite{soppina2022kinesin}. In the panels, the blue curves in (a) and (b) correspond to analytical expression $\langle W_c\rangle $ in Eq.\ref{eq:Wc1} and for (c) and (d) it corresponds to analytical expression of $P(W_c)$ in Eq.\ref{Pw}. Comparison is done with 2D stochastic simulation with bead radius $R = 0.05 ~ \mu m$(red squares) and $R = 0.1 ~ \mu m$ (green circles). For simulations, motor rest length $l_o = 110 ~ nm$ \cite{erickson2011molecular,uccar2019force} and trap stiffness along the MT and perpendicular to it are chosen as $k_o^x=k_o=0.005~pN~nm^{-1}$ and $k_o^y=k_o^x/3$ \cite{ashkin1992biophysj,bormuth2008opticsexp}, respectively. All other parameters are same as Fig.\ref{fig3} for all panels. Stochastic simulation curves are obtained from $10^6$ independent runs.}
    
    %\caption{Engine characteristics: Effect of variation of motor velocity (at zero load) : (a) Average Work done by motor per cycle: $(W_m)$ vs $v_o$, (b) Average power output: $P$ vs $v_o$ and (c) Efficiency defined as the ratio of the average work output of the system and average work done by motor: $\eta$ vs $v_o$. Here $k_o = 0.005 ~ pN ~nm^{-1}$. For {\it kinesin-1} motor, $\epsilon_o = 0.72 ~ s^{-1}$, $f_s = 5.7 ~$pN, and $f_m = 4$ pN \cite{brenner2020force}, while for  {\it kinesin-3} motor, $\epsilon_o = 0.23 s^{-1}$, $f_s = 3$pN, and $f_m = 2.7$ pN \cite{budaitis2021pathogenic,soppina2022kinesin}.}
    \label{fig5}
\end{figure}

\emph{2D Stochastic model} --- 
While for the 1D model, bead movement only along the MT axis is taken into account, in general, the vertical distance between the center of mass of the bead and MT changes during the engine cycle (Fig.S7). We perform a stochastic simulation that accounts for the full 2D movement of the bead along with the effects of finite rest length of the motor and its stochastic stepping on MT with discrete step size (see section VI of SM for details).
%We determine the average work per cycle and the PDF of work both for the case of \textit{kinesin-1} and \textit{kinesin-3} motors . We find that $\langle W_c \rangle $ and $P(W_c)$ for engines driven by both \textit{kinesin-1} and \textit{kinesin-3}, there is fairly good agreement between the 2D stochastic simulations performed with sub-micron sized bead and the analytical model predictions (Fig.\ref{fig5}).
We find that for sub-micron size beads, comparison of $\langle W_c \rangle$ and $P(W_c)$ for engines driven by \textit{kinesin-1} and \textit{kinesin-3} show fairly good agreement with the analytical 1D model predictions (Fig.\ref{fig5}).

\emph{Effect of time delay in feedback} --- So far we have considered an instantaneous feedback mechanism of the trap stiffness without any delay. A time delay $\delta t$ in implementing the feedback process of changing the trap stiffness $k_t$ would result in loss of work output. While the effect of time delay at attachment step of the motor results in a reduction of work output per cycle by an amount $
\delta W_{c}^a \simeq \delta t_a \left( \frac{\mu v_o^{2}}{2 \epsilon_o^2} \right)$, the reduction of work output per cycle due to time delay in the detachment step of the motor is $\delta W_{c}^{d} \simeq \delta t_d \left[ \mu v_o^{2} \langle \tau_1^{3}\rangle \left(\frac{k_o}{\gamma}\right)\right]$ (see section VII of SM for details.). Thus, for time delay beyond typical relaxation time for the Brownian particle in the harmonic trap ($\tau_b=\gamma/k_o$), no useful work can be extracted from the engine. Indeed if $\delta t_d$ exceeds $\tau_b$, then the bead would have already relaxed to the trap center, and consequently no work output can be obtained.
%therefore the effect of changing the trap stiffness would not result in work output. 
%and it sets a bound for the performance of the engine. 

\emph{Experimental feasibility} --- Experimental realization of this microengine is contingent on the requirement $\delta t_d \ll \gamma/k_o$. Since the change of the trap stiffness in the optical trap is achieved by modulating the laser power, the time-delay in switching the laser power back to its initial value should be much less than the time needed for the bead to relax to the trap center. The typical displacement of the bead from the trap center is $\sim~ 400~ nm$.  %According to the proposal of this microengine, 
%Thus the time-delay in switching the laser power back to its initial value should be much less than the time needed for the bead to relax to the trap center. 
%Assuming $k_o=5 \times 10^{-3}~pN nm^{-1} $, the lower limit on trap stiffness, the thermal fluctuations in the bead position in a force balanced condition is $30 ~nm$. %This is obtained by equating the mean square displacement to the thermal energy  divided by $k_0$. 
 Once the bead moves more than $~8 ~nm$ (stochasticity associated with motor stepping size) {\it towards} the trap center,  a Transition-Transition Logic (TTL) can trigger to  switch the laser power to modulate the trap stiffness. %back to its initial value,  completing the cycle.  
%In both these steps laser power can be directly controlled by current to the laser. 
Given that the laser intensity can be modulated at 100 KHz by direct modulation of current, the delay arising out of lasing the cavity at this new power would be smaller than $10 ~\mu s$, roughly 10 times less than the time required for a sub-micron size bead to relax to the trap center ($\gamma/ k_o\sim 10^{-4} s$). Thus, achieving the condition of the time delay of $\delta t_d \ll \gamma/k_o$ using current Infrared Lasers is very much feasible.

\emph{Discussion} ---  To summarize, we have conceptualized a unique  work-to-work converter microengine and provided concrete quantitative predictions of engine performance for feasible biological parameters of motor proteins. This engine is capable of extracting work much larger than other micro-machines reported so far.
%In this article, we have conceptualized a unique  work-to-work converter microengine which is capable of extracting work much larger than other micro-machines reported so far.  
%in a conventional optical trap set up. This engine is powered by a single motor protein and functions as a work-to-work converter for colossal work extraction (in comparison to other micro-machines reported so far), harnessing the motility of the motor protein into the cyclic work output of the engine. 
A distinctive feature of this engine which delineates it from other engines is that the fidelity and performance of this engine are determined by the stochasticity of the motor (un)binding process and not by the (a)thermal bath characteristics.
%The distinctive feature which delineates this engine from other microengines, is that the {\it fidelity} of the engine is determined by the stochasticity of the motor (un)binding process in contrast to other microengines where it is determined by the noise characteristics of the thermal / athermal bath in which the engine operates. 
Our analytical calculations of the average work and its PDF, which are validated by stochastic simulations, set the ground for realization of a much more powerful microscale engine which is a promising prototype for fabricating a microscale device in future.

This study also highlights the role of information in shaping the performance of engine. 
%Indeed the true thermodynamic cost of such information engine has to account for the information entropy associated for error free measurement 
%It also sets the stage for investigating the intimate connection of thermodynamics and Information for driven system.
In particular, it may be noted that the information content for this system is associated with the stochasticity (uncertainty) of motor (un)binding process from MT and motor stepping on MT. Information entropy for error free measurements \cite{shannon1998mathematical} and erroneous measurements \cite{thomas2006elements} of motor position and its states (bound or unbound) will contribute to the true thermodynamic cost of this machine. 
%Together with other costs (e.g. energetic cost from ATP hydrolysis) these thermodynamic costs should also contribute to the true efficiency of the machine. 
Formulating and calculating the information estimates for this motor-powered engine would set the stage for investigating the intimate connection between thermodynamics and information for driven systems in general and motor protein-driven engines in particular. Work along this direction is in progress. 
Another possible future direction to explore would be to come up with a design of a microengine that is powered by multiple motors.  
%While in such scenario, the average work output per cycle is expected to increase, the stochasticity associated with multiple motors (un)binding would adversely affect the fidelity of the engine. 
%Another facet which can be worthwhile to explore is whether other types of motor proteins can be used as a working substance. 
Interestingly, dynein motors are {\it catchbonded}, exhibiting increased lifetime of bond to MT under load force \cite{sundararajan2024theoretical,guha2021novel}. It remains to be determined whether the effect of {\it catchbonding} would lead to improvement of engine performance.

\emph{Acknowledgements} --- S.M. acknowledges financial support from University of Hyderabad under the startup grant (OH-35-CA-2024-25) and grant from ANRF-Pair program (ANRF/PAIR/2025/000012/EPAIR). S.M. also acknowledges financial support and hospitality for visit to ICTP, Trieste under the Associateship program, where part of the work was done. A.S. acknowledges the Core Research Grant (CRG/2019/001492) from DST, Government of India. A.S. also acknowledges funding by the CY Initiative of Excellence (grant “Investissements d’Avenir” ANR-16-IDEX-0008) and the work was partially developed during his stay at CY Advanced Studies whose support is gratefully acknowledged.

%\section*{Author Contributions}
\emph{Author Contributions} --- S.M designed the study. S.M, A.S, S.D conceived the working principle. S.M, S.D  and S.G performed the analytic calculations. S.D and B.R performed numerical integrations. S.G, S.D and B.R performed stochastic simulations. S.M, A.S, S.D, S.G analyzed the data. S.P provided inputs for experimental feasibility. All authors reviewed and wrote the manuscript.

%\section*{Competing Interests}
\emph{Competing Interests} --- The authors declare no conflict of interest.

%\authorcontributions{S.M designed the study. S.M, A.S, S.D conceived the working principle.  S.M and S.D performed the analytic calculations. S.D, B.R performed numerical integrations. S.G, B.D performed stochastic simulations. S.M, A.S, S.D, S.G analyzed the data. S.P provided inputs for experimental feasibility. All authors reviewed and wrote the manuscript.}

%\authordeclaration{The authors declare no conflict of interest.}

%\showacknow{} % Display the acknowledgments section

%\renewcommand{\figurename}{FIG. S}
\renewcommand{\thefigure}{S\arabic{figure}}
\setcounter{figure}{0} 

\section{Appendix A: Expression for averaged Position of bead}

When the motor is bound to the Microtubule (MT) filament, evolution of the bead is governed by the overdamped Langevin Equation,
\begin{equation}
    \gamma\dot{x} = -k_t(t)x +f(t)+\xi(t)
    \label{langevinEQ}
\end{equation}
 Here, the dynamics of bead is coupled to that of the motor through the force $f(t)= k_m[x_m(t)-x(t)]$ which is a function of both- motor as well as the bead variables. The motor position evolves according to,
 \begin{equation}
     \dot{x}_m = v_o(1-f/f_s)
     \label{linearv}
 \end{equation}
These two equations can be decoupled in the Laplace domain. To do so, first we average out the thermal noise $\xi(t)$ which has zero mean $\overline{\xi}(t) = 0$ and is delta correlated $\overline{\xi(t)\xi(t')} = 2\gamma k_BT\delta(t-t')$. We then take the Laplace transform of Eq.\ref{langevinEQ} resulting in following expression-
\begin{equation}
    \gamma s \overline{x}(s) = -(k_o + k_m)\overline{x}(s) + \mu\frac{d}{ds}\overline{x}(s) + k_mx_m(s)
    \label{laplace-langevin}
\end{equation}
In the same way Laplace transform of Eq.\ref{linearv} is,
\begin{equation}
    x_m(s) = \frac{f_s}{sf_s + v_ok_m}\left[\frac{v_o}{s} + \frac{v_ok_m}{f_s}\overline{x}(s)\right]
     \label{laplace-vm}
\end{equation} 
We can now substitute $x_m(s)$ (Eq.\ref{laplace-vm}) in $\overline{x}(s)$ (Eq.\ref{laplace-langevin}) thereby eliminating the motor variables from the equation of motion of the bead. We also define the quantities $q = k_mv_o/f_s$ and $\alpha=qk_o/(k_o + k_m)$ to express it as,

\begin{equation}
    \left(\tau_bs - \frac{1}{\tau_p}\frac{d}{ds}\right)\overline{x}(s) = \frac{k_o + k_m}{k_o}\left[\frac{f_s\alpha}{k_os(s+q)}-\left(\frac{s+\alpha}{s + q}\right)\overline{x}(s) \right]
    \label{fullLaplace}
\end{equation}
% \begin{equation}
%     \left(\tau_bs - \frac{1}{\tau_p}\frac{d}{ds}\right)\overline{x}(s) = -\frac{k_o + k_m}{k_o}\left[\left(\frac{s+\alpha}{s + q}\right)\overline{x}(s) \nonumber\\
%     - \frac{f_s\alpha}{k_os(s+q)}\right]
%     \label{fullLaplace}
% \end{equation}

Here, $\tau_b = \gamma/k_o$ is the hydrodynamic relaxation time scale, representing the time required for the bead to relax back to the trap center. For our system, with bead-size $R\sim~0.1 ~\mu m$ and optical trap stiffness $k_o$ in the range $(0.005- 0.2) ~pN ~nm^{-1}$ this is a fast process- $\tau_b\sim {10}^{-4} ~s$ relative to the timescales $(10^{-3} -10^{-1}) ~s$ over which motor takes its steps. On the other hand $\tau_p = k_o/\mu$ is the time scale over which the trap stiffness $k_o$ is tuned. For quasi-static (or slow) protocols $\tau_p$ is larger than timescales associated with motor stepping. In our simulations we take $\mu = 0.1k_o ~pN ~nm^{-1}~s^{-1}$, so that $\tau_p \sim 10 ~ s$. 
As the timescale associated with motor stepping is intermediate to $\tau_b$ and $\tau_p$, we take the limits, $\tau_b \rightarrow 0$ and $\tau_p \rightarrow \infty$ making left hand side of Eq.\ref{fullLaplace} zero, allowing the expression to be rearranged as,
\begin{equation}
    \overline{x}(s) = \frac{f_s}{k_o}\frac{\alpha}{s(s+\alpha)}
    \label{laplace-xc}
\end{equation}
The inverse Laplace transform of above expression gives,
\begin{equation}
   \overline{x}(t) = \frac{f_s}{k_o}(1- e^{-\alpha t})
   \label{Eq:bead-av}
\end{equation}

\begin{figure}[h]
    \centering
    \includegraphics[width=\linewidth]{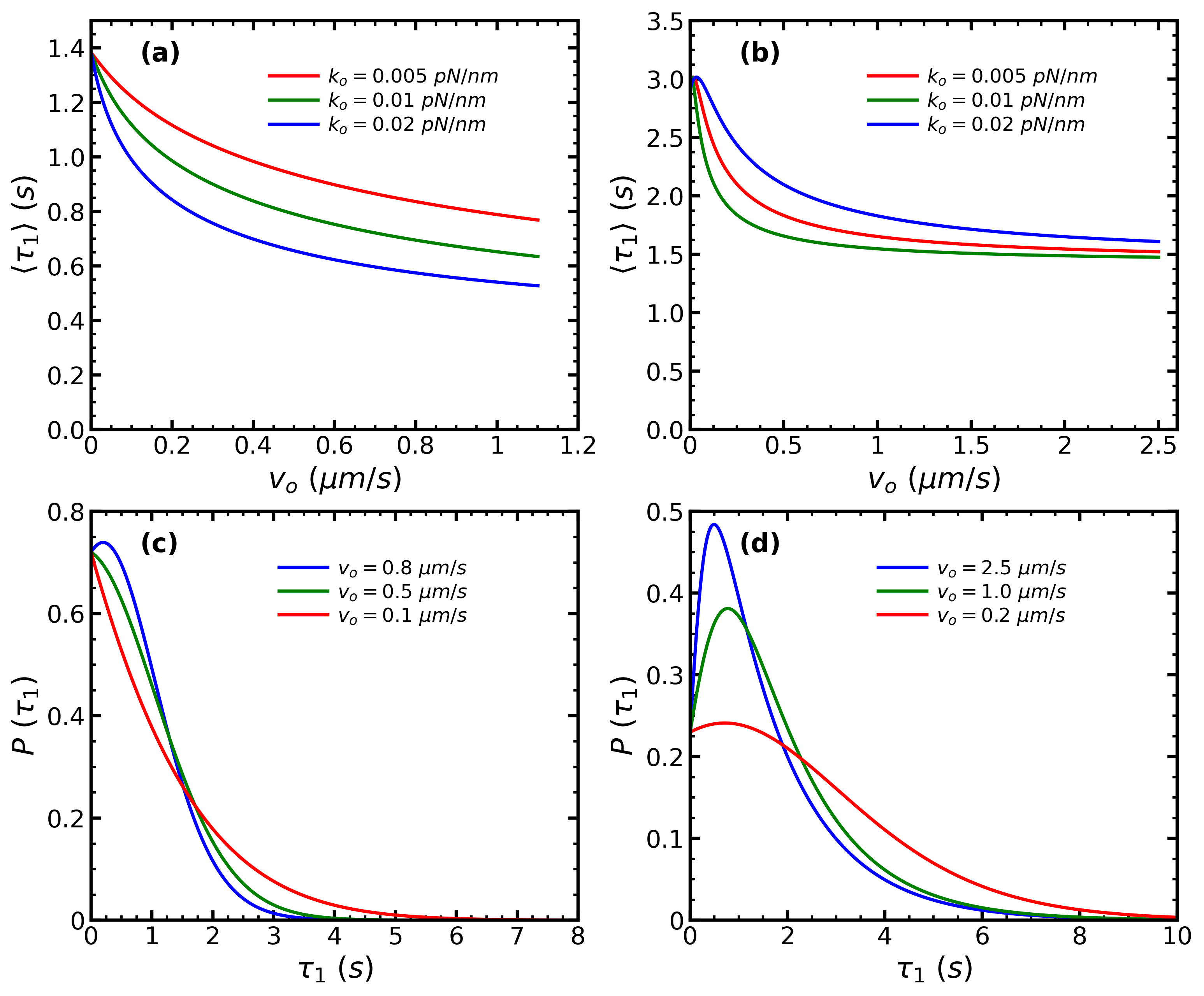}
    \caption{ Mean run time $\langle \tau_1\rangle$ for (a) \textit{kinesin-1} (b) \textit{kinesin-3} as a function of $v_o$. Run time distribution Eq.\ref{eq:P1} $P(\tau_1)$ (c) for \textit{kinesin-1}. Here, $k_o = 0.005 ~ pN~ nm^{-1}$, $f_s = 5.7~ pN$ , $f_m = 4.0 ~pN$ ,  $\epsilon_o = 0.72 ~ s^{-1}$ .  
    (d) for \textit{kinesin-3}. Here, $k_o = 0.005 ~ pN~ nm^{-1}$, $f_s=3.0 ~ pN$ , $f_m = 2.7 ~ pN$ , $\epsilon_o = 0.23 ~ s^{-1}$.
    }
    %all curves exhibit a peak as for all choices, velocity exceeds the critical value of $0.12 ~ \mu m/s$.
    %A peak is visible for $v_o = 0.8 ~ \mu m/s$ (blue curve) as velocity exceeds the critical value of $ 0.57 ~\mu m/s$. 
    \label{fpt}
\end{figure}

\begin{table}[h!]
\label{tab:Table}
\begin{center}
\begin{tabular}{ |c|c|c| }
\hline
%\rowcolor{Gray}
Parameter                           & Symbol       & Value  \\
\hline
Binding rate                        & $\pi_o$      & 1 $s^{-1}$   \cite{leduc2004cooperative,klumpp2005cooperative}\\ \hline
Unbinding rate     & $\epsilon_o$ & 0.1 -- 1.0 $s^{-1}$  \cite{soppina2022kinesin,klumpp2005cooperative}\\ \hline
Principal velocity & $v_o$        & 30 -- 3000 $ nm$ $s^{-1}$ \cite{howard1989movement,soppina2022kinesin}\\ \hline
Motor rest length             & $l_o$        & 110 $nm$  \cite{erickson2011molecular,uccar2019force} \\ \hline
Stall force              & $f_s$        & 3 - 6 $pN$  \cite{budaitis2021pathogenic, brenner2020force}\\ \hline
Detachment force         & $f_m$        & 2 - 4 $pN$  \cite{guo2019force, budaitis2021pathogenic}\\ \hline
%Catch bond threshold force          & $f_m$        & chosen same as $f_s$  \cite{puri2019prr} \\ \hline
%Deformation force scale             & $f_o$        & 7 $pN$  \cite{puri2019prr} \\ \hline
%Catch bond Strength / Deformation energy scale             & $\alpha$        & 0 - 40 $k_B T$          \cite{puri2019prr} \\ \hline
Motor spring stiffness              & $k_m$        & 0.3 $pN$ $nm^{-1}$   \cite{coppin1997load}\\ \hline
Trap Stiffness                      & $k_o$        & 0.005 -- 0.03 $pN$ $nm^{-1}$  \cite{rai2013molecular,brenner2020force}\\ \hline
kinesin step size & $d$ & $8~nm$ \cite{schnitzer1997nature} \\
\hline
\end{tabular}
\end{center}
\caption{Experimental values of physical parameters for kinesin motor proteins and optical trap.}
  \label{table1-sup}
\end{table}

\section{Appendix B: Characteristics of Run Time}

The probability distribution of run time (or first passage time) takes the form,
\begin{equation}
    P(\tau_1)=  \epsilon_{o} \exp \left[ \frac{f_s}{f_m} \left( 1 - e ^{- \alpha \tau_1} \right )\right ]  \exp \left [ -\int_{0}^{\tau_1} \epsilon(t) dt\right ]
\label{eq:P1}
\end{equation}
where the explicit functional form of the unbinding rate as function of time is 
\begin{equation}
 \epsilon (t) = \epsilon_{o} \exp \left[ \frac{f_s}{f_m} \left( 1 - e ^{- \alpha t}\right) \right]
\end{equation}

The mean run time for \textit{ kinesin-1} and \textit{kinesin-3} is plotted in   Fig.\ref{fpt}(a) and (b) respectively. Fig.\ref{fpt}(c) and (d) depict the run time distribution $P(\tau_1)$ at different velocities for a optical trap stiffness of $k_o = 0.005 ~pN~ nm^{-1}$. When $\alpha < \alpha_c$ the run time distribution $P(\tau_1)$ is monotonically decreasing and exhibits a peak once $\alpha > \alpha_c$ where the critical value $\alpha_c = \epsilon_of_m/f_s$. For a trap stiffness of $k_o = 0.005 ~pN~ nm^{-1}$, the run time distribution exhibits a peak if the velocity $v_o > 0.57 ~\mu m~ s^{-1}$ for \textit{kinesin-1} while $v_o>0.12 ~ \mu m~ s^{-1}$ for \textit{kinesin-3}.

The mode of run time distribution ( i.e. the value of $\tau_1$ that maximizes $P(\tau_1)$) can be obtained by equating the first derivative to zero. The result, is expressible in terms of the Lamberts-$W$ function- $W_k(x)$. For the range of physical parameters (see Table.\ref{table1-sup}) the argument of Lamberts function is always real and positive, in which case there is only one real solution given by the principle branch $k=0$.
\begin{equation}\label{eq-taupeak}
    t_o(\alpha) = \frac{1}{\alpha}\left[\ln\left(\frac{f_s\alpha}{f_m\epsilon_oe^\frac{f_s}{f_m}}\right) + W_o\left(\frac{e^{\frac{f_s}{f_m}}\epsilon_o}{\alpha}\right)\right]
\end{equation}

\begin{figure}[h!]
    \centering
    \centering
    \includegraphics[width=0.9\linewidth]{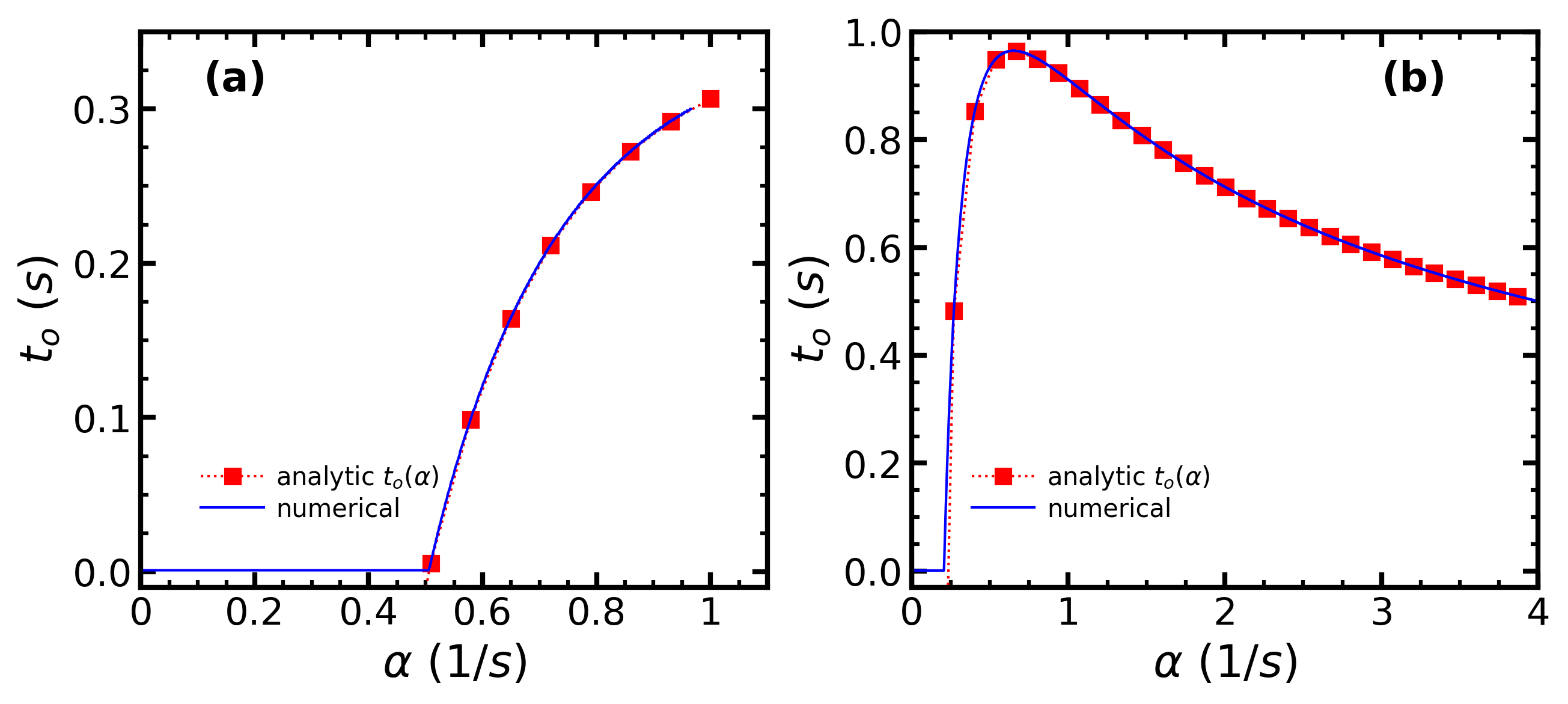}
    \caption{Comparison of numerically estimated mode (blue curve) of $P(\tau_1)$ with the analytic expression Eq.\ref{eq-taupeak} (red squares). As long as $\alpha < \alpha_c$, the distribution is monotonic decreasing and the maxima happens to be at zero. Once $\alpha>\alpha_c$ mode becomes non-zero. (a) For \emph{kinesin-1} motor, $\alpha_c = 0.505 ~s^{-1}$. Here $f_s = 5.7~ pN$ , $f_m = 4.0 ~pN$ ,  $\epsilon_o = 0.72 ~ s^{-1}$ . (b) For \emph{kinesin-3} motor, $\alpha_c = 0.207 ~s^{-1}$. Here $f_s=3.0 ~ pN$ , $f_m = 2.7 ~ pN$ , $\epsilon_o = 0.23 ~ s^{-1}$.}
    \label{tauplot}
\end{figure}

The run time distribution exhibits a peak only when $\alpha > \alpha_c$. For all other values $\alpha \leq \alpha_c$ we have $t_0 \leq 0$ where the equality holds only if $\alpha = \alpha_c$. Fig.\ref{tauplot} compares Eq.\ref{eq-taupeak} with the numerically estimated mode of $P(\tau_1)$.

\section{Appendix C: Calculation and comparison between $\overline{x^2}$ and $\overline{x}^2$}
The overdamped Langevin equation can be written as,
\begin{eqnarray}
    \gamma\dot{x}&=&-k_t(t)x+f(t)+\xi(t) \nonumber\\
    \Rightarrow \gamma\dot{x}&=&-(k_o+\mu t)x+k_m(x_m-x)+\xi(t)\nonumber\\
    \Rightarrow \frac{dx}{dt}&+&\frac{(k_o+\mu t+k_m)}{\gamma}x=\frac{k_mx_m}{\gamma}+\frac{\xi(t)}{\gamma}\nonumber
\end{eqnarray}
The integrating factor in this case becomes, $I(t)=e^{kt/\gamma}$ where $k=k_o+\frac{1}{2}\mu t+k_m$. Therefore, the solution of Langevin equation can be written as,
\begin{eqnarray}
    %& &e^{kt/\gamma} \cdot\frac{dx}{dt}+e^{kt/\gamma}\cdot\frac{(k_o+\mu t+k_m)}{\gamma}x=\frac{e^{kt/\gamma}}{\gamma}\left[k_mx_m+\xi(t)\right] \nonumber\\
    &\Rightarrow& \frac{d}{dt}\left(xe^{kt/\gamma}\right)=\frac{e^{kt/\gamma}}{\gamma}\left[k_mx_m+\xi(t)\right] \nonumber\\
    &\Rightarrow& \int_0^{t}\frac{d}{dt}\left(xe^{kt/\gamma}\right)dt=\int_0^{t}\frac{e^{kt/\gamma}}{\gamma}\left[k_mx_m+\xi(t)\right] dt \nonumber
\end{eqnarray}
As the bead always starts from the trap center, $x(0)=0$. Therefore,
\begin{eqnarray}
    & &x=e^{-kt/\gamma}\int_0^{t} \frac{e^{kt/\gamma}}{\gamma}\left[k_mx_m+\xi(t)\right] dt \nonumber\\
    &\Rightarrow& \overline{x}=e^{-kt/\gamma}\int_0^{t} \frac{e^{kt/\gamma}}{\gamma}k_mx_m dt ~~~[\because \overline{\xi}(t)=0]
\end{eqnarray}
Now,
\begin{eqnarray}
    & &x^2=e^{-2kt/\gamma}\int_0^{t} \int_0^{t} \frac{e^{k(t_1+t_2)/\gamma}}{\gamma^2}\left[k_mx_m(t_1)+\xi(t_1)\right]\nonumber\\&~&~~~~~~~~~~~~~~~~~~~~~~~~~~~~\times\left[k_mx_m(t_2)+\xi(t_2)\right] dt_1 dt_2 \nonumber
    \end{eqnarray}
    %&\Rightarrow& x^2=e^{-2kt/\gamma}\int_0^{t} \int_0^{t} \frac{e^{k(t_1+t_2)/\gamma}}{\gamma^2}\left[k_m^2x_m(t_1)x_m(t_2)+k_m(x_m(t_1)\xi(t_2)+x_m(t_2)\xi(t_1)+\xi(t_1)\xi(t_2)\right]dt_1dt_2 \nonumber\\
    %&\Rightarrow& \overline{x^2}=e^{-2kt/\gamma}\int_0^{t} \int_0^{t} \frac{e^{k(t_1+t_2)/\gamma}}{\gamma^2}\left[k_m^2x_m(t_1)x_m(t_2)+2\gamma k_bT\delta (t_1-t_2)\right]dt_1dt_2 \nonumber\\
%    &\Rightarrow& \overline{x^2}=\overline{x}^2+e^{-2kt/\gamma}\int_0^{t} \int_0^{t} \frac{e^{k(t_1+t_2)/\gamma}}{\gamma}\cdot 2k_BT \delta(t_1-t_2) ~dt_1dt_2 \nonumber\\

It then follows that, 
\begin{equation}
\overline{x^2}=\overline{x}^2+e^{-2kt/\gamma}\int_0^{t}\frac{e^{2kt_1/\gamma}}{\gamma}\cdot 2k_BT ~dt_1 
    %&\Rightarrow& \overline{x^2}=\overline{x}^2+k_bT\sqrt{\frac{\pi}{\gamma\mu}}e^{-2k\tau_1/\gamma}e^{-\frac{(k_0+k_m)^2}{\gamma\mu}}\left[\text{erfi}\left(\frac{k_m+k_0+\mu \tau_1}{\sqrt{\gamma\mu}}\right)-\text{erfi}\left(\frac{k_m+k_0}{\sqrt{\gamma\mu}}\right) \right] \nonumber\\
    \label{eq:x2bar}
\end{equation}
The integral in Eq.\ref{eq:x2bar} has an exact solution in terms of imaginary error function.
% Notice, we can arrage the srguments in square bracket as,
% \begin{eqnarray}
%     & &\text{erfi}\left(\frac{k_m+k_0}{\sqrt{\gamma\mu}} + \frac{\mu \tau_1}{\sqrt{\gamma\mu}}\right)-\text{erfi}\left(\frac{k_m+k_0}{\sqrt{\gamma\mu}}\right)\\
%     &\Rightarrow&
%     \text{erfi}\left(\frac{k_m+k_0}{\sqrt{\gamma\mu}} + \frac{\tau_1}{\sqrt{\tau_b\tau_p}}\right)-\text{erfi}\left(\frac{k_m+k_0}{\sqrt{\gamma\mu}}\right)
% \end{eqnarray}
But as for our system, $t\leq \tau_1 \sim \mathcal{O}(1)$ for both \emph{kinesin-1} and \emph{kinesin-3}, and $\mu t< k_o \ll k_m$ by choice, one can safely assume $k=(k_o+\mu t+k_m)\approx k_m$. Hence Eq.\ref{eq:x2bar} becomes,
\begin{eqnarray}
\overline{x^2} &\simeq& \overline{x}^2+e^{-2k_mt/\gamma}\int_0^t \frac{e^{2k_mt_1/\gamma}}{\gamma} \cdot 2k_BT~dt_1 \nonumber\\
\overline{x^2} &\simeq& \overline{x}^2+\frac{k_BT}{k_m}\left[1-e^{-2k_mt/\gamma}\right]\nonumber\\
\overline{x^2} &\simeq& \overline{x}^2+\frac{k_BT}{k_m} ~~~~\left[\because e^{-2k_mt/\gamma}\rightarrow0~~\text{as}~~ k_mt/\gamma \sim 10^7\right] \nonumber
\end{eqnarray}
Therefore, the typical work output for the system would be,
\begin{eqnarray}
    \Delta W_c^{(AB)} &=& \frac{\mu}{2}\int_0^{\tau_1}\overline{x^2}dt \nonumber\\
    &\simeq& \frac{\mu}{2} \int_0^{\tau_1} \overline{x}^2~dt+ \int_0^{\tau_1}\frac{\mu k_BT}{k_m}~dt \nonumber\\
    &\simeq& \frac{\mu}{2} \int_0^{\tau_1} \overline{x}^2~dt+ \frac{\mu \tau_1 k_BT}{k_m} \nonumber
\end{eqnarray}
As for our system, $\tau_1\sim \mathcal{O}(1)$, $\mu=k_o/10=0.0005~pN~nm^{-1}~s^{-1}$, and $k_m=0.3~pN~nm^{-1}$, we have $\frac{\mu \tau_1}{k_m}\sim 10^{-3}$ and hence the work output becomes,
\begin{equation}
    \Delta W_c^{(AB)}\simeq  \frac{\mu}{2} \int_0^{\tau_1} \overline{x}^2~dt+ 10^{-3}k_BT
\end{equation}
Similarly $\Delta W_c^{(BC)}$ can be written as,
\begin{eqnarray}
        \Delta W_c^{(BC)}&=&-\frac{1}{2}\mu \tau_1 \overline{x^2}(\tau_1)\nonumber\\
        &\simeq& -\frac{1}{2}\mu \tau_1 \overline{x}^2(\tau_1)-10^{-3}k_BT
\end{eqnarray}
Also the work done by the motor can be calculated as,
\begin{eqnarray}
            \Delta W_m^{(AB)}&=&\int_0^{\tau_1}f(t)\dot{x}dt=\int_0^{\tau_1}k_m(x_m-x)\dot{x}dt \nonumber\\
        &=&k_m\left[x_mx(\tau_1)-\frac{x^2(\tau_1)}{2}\right]
\end{eqnarray}
Now taking the thermal average one can easily obtain,
\begin{eqnarray}
    \Delta W_m^{(AB)}&=&k_m\left[x_m\overline{x}(\tau_1)-\frac{\overline{x^2}(\tau_1)}{2}\right] \nonumber\\
    &\simeq& k_m\left[x_m\overline{x}(\tau_1)-\frac{\overline{x}^2(\tau_1)}{2}\right] -k_BT
\end{eqnarray}
Given that in our system, for typical motor parameters the average work output $\Delta W_c\sim10~k_BT$ and input $\Delta W_m\sim50-200~k_BT$ (shown later), the choice of calculating work output using $\overline{x}^2$, instead of $\overline{x^2}$, does not significantly affect our results.
% Therefore, the (thermally averaged) potential energy of the bead due to the optical trap can be written as,
% \begin{equation*}
%     \overline{U}(t)=\frac{1}{2}k_t(t)\overline{x^2}=\frac{1}{2}k_t(t)\left[\overline{x}^2+\frac{k_bT}{k_m}\right]
% \end{equation*}
% Now $k_t(t)\sim k_o$ as $\mu=k_o/10$ and $t\leq\tau_1\sim\mathcal{O}(1)$, we have
% \begin{equation*}
%     \frac{k_t(t)}{k_m}\simeq \frac{k_o}{k_m} \sim 10^{-2} - 10^{-1}
% \end{equation*}
% Therefore, finally we have,
% \begin{equation}
%     \overline{U}(t)=\frac{1}{2}k_t(t)\overline{x^2}\simeq\frac{1}{2}k_t(t)\left[\overline{x}^2+0.1k_bT\right]\simeq \frac{1}{2}k_t(t)\overline{x}^2
% \end{equation}
% \textbf{SG : It is worthwhile to note that $\overline{x^2}\geq\overline{x}^2$ and hence all average work output predictions in this manuscript is representing a lower bound of the actual work output $\langle W_c \rangle$.}}

\section{Appendix D: Stochastic thermodynamics of the engine}

By integrating over single trajectories, the Langevin equation (Eq.\ref{langevinEQ}) can be cast into the First law of thermodynamics which has the generic form $ \Delta U = \Delta Q + \Delta W_c + \Delta W_{m}$ where $\Delta U$ is the internal energy change,  $\Delta Q$ is the heat input, $\Delta W_c$ is the work output of the engine, and $\Delta W_{m}$ is the work done by the motor protein when it is attached to MT.  For our system, the internal energy $U(x,k_t)=\frac{1}{2}k_t(t)x^2$. It follows that the corresponding expression for $ \Delta Q = -\int [\gamma \dot{x}(t) - \xi(t)]\dot{x}(t)dt$, $\Delta W_c =\int\frac{\partial U}{\partial k_t}\dot{k}_tdt$ and $\Delta W_{m} = \int f(t)\dot{x}(t)dt$. Next, we obtain the thermally averaged expression for work and heat in each step of the engine cycle. 

%\textcolor{blue}{SD: We have not mentioned sign convention for thermodynamic quantities. For instance, " a negative sign for energy corresponds to energy going out of system while positive sign corresponds to energy coming into the system." }

\subsection{Expression for Thermodynamic quantities related to engine}
%By integrating over single trajectories, the Langevin equation (Eq.\ref{langevinEQ}) can be cast into the First law of thermodynamics which has the generic form $ \Delta U = \Delta Q + \Delta W_c + \Delta W_{m}$ where $\Delta U$ is the internal energy change,  $\Delta Q$ is the heat input, $\Delta W_c$ is the work output of the engine, and $\Delta W_{m}$ is the work done by the motor protein when it is attached to MT.  For our system, the internal energy $U(x,k_t)=\frac{1}{2}k_t(t)x^2$. It follows that the corresponding expression for $ \Delta Q = -\int [\gamma \dot{x}(t) - \xi(t)]\dot{x}(t)dt$, $\Delta W_c =\int\frac{\partial U}{\partial k_t}\dot{k}_tdt$ and $\Delta W_{m} = \int f(t)\dot{x}(t)dt$. We next calculate the thermally averaged expression of work and heat done in an engine cycle. 
\vskip 0.2 cm

\textit{Step AB:} In this step, the trap stiffness increases linearly with time until the motor detaches at $\tau_1$. Since $\overline x^2 \simeq \overline{x^2}$, it follows that, 
\begin{equation}
\Delta W^{(AB)}_{c} = \int_{o}^{\tau_1}\left( \frac{\partial \overline{U}}{\partial k_t}\right) \dot {k}_t dt = \frac{\mu}{2} \int_{o}^{\tau_1} {\overline x }^{2} dt
\label{eq-AB}
\end{equation}
Note that, here work is still a stochastic quantity, since $\tau_1$ is a stochastic quantity. The corresponding expression for work done by the motor is,
\begin{equation}
\Delta W_{m}^{(AB)} = \int_0^{\tau_1} f(t)\dot{\overline x}(t)dt
\end{equation}
Using Eq.\ref{Eq:bead-av} and force-balance condition we obtain $\Delta W_m^{(AB)} = - \frac{1}{2} \frac{k_o v_{o}^{2}}{\alpha^2} \left(1 + e^{-2\alpha\tau_1} - 2 e^{-\alpha\tau_1}\right)$. 
The expression of internal energy change is,
\begin{equation}
\Delta U^{(AB)} = \frac{1}{2}(k_o + \mu \tau_1) {\overline x}^2(\tau_1)
\end{equation}
The expression for the heat input $\Delta Q^{(AB)}$ can then be obtained by using the form of First Law of Thermodynamics. %$\Delta Q^{(AB)} = \Delta \overline{U}^{(AB)} - \Delta W_c^{(AB)} -\Delta W_{m}^{(AB)}$.

\vskip 0.2 cm
\textit{Step BC}: At time $\tau_1$ the motor detaches from MT and the trap stiffness is instantaneously changed from $k_o + \mu \tau_1$ to $k_o$. During this instantaneous jump process $\Delta Q^{(BC)} =0$ and as the motor remains inactive the work done by the motor $\Delta W^{(BC)}_m =0$. Therefore, work output of the engine is simply the change in internal energy, 
\begin{equation}
 \Delta W^{(BC)}_c= \Delta {U}^{(BC)} = -\frac{1}{2}\mu \tau_1 {\overline x}^{2}(\tau_1) 
 \label{eq-BC1}
 \end{equation}
\\

\textit{Step CA:} In this step, the bead relaxes back to trap center $\overline{x} = 0$ and the internal energy reduces to zero. Neither the system does any work $\Delta W_c^{(CA)}=0$, nor the motor, since it has already detached from MT filament, so that $\Delta W_m^{(CA)}=0$. The lower internal energy is achieved by dissipating heat into the environment. Thus it follows that 
\begin{equation}
\Delta {U}^{(CA)} = \Delta {Q}^{(CA)} =-\frac{1}{2}k_o {\overline x}^{2}(\tau_1) 
\end{equation}

We note that for the entire cycle, the work input due to the work done by the motor on the system gets converted into work output by the system and the difference is dissipated as heat. 

The total work output in one entire cycle of the engine is obtained by summing up the contributions from all the steps. It then follows that  
\begin{eqnarray}
     W_c &=& \Delta W_c^{(AB)} + \Delta W_c^{(BC)} + \Delta W_c^{(CA)}\nonumber\\
     &=& \frac{\mu}{2} \int_{o}^{\tau_1} {\overline x }^{2} dt-\frac{1}{2}\mu \tau_1 {\overline x}^{2}(\tau_1) \label{W_c} 
\end{eqnarray}

\begin{figure}[t]
    \centering
    \includegraphics[width=0.5\linewidth]{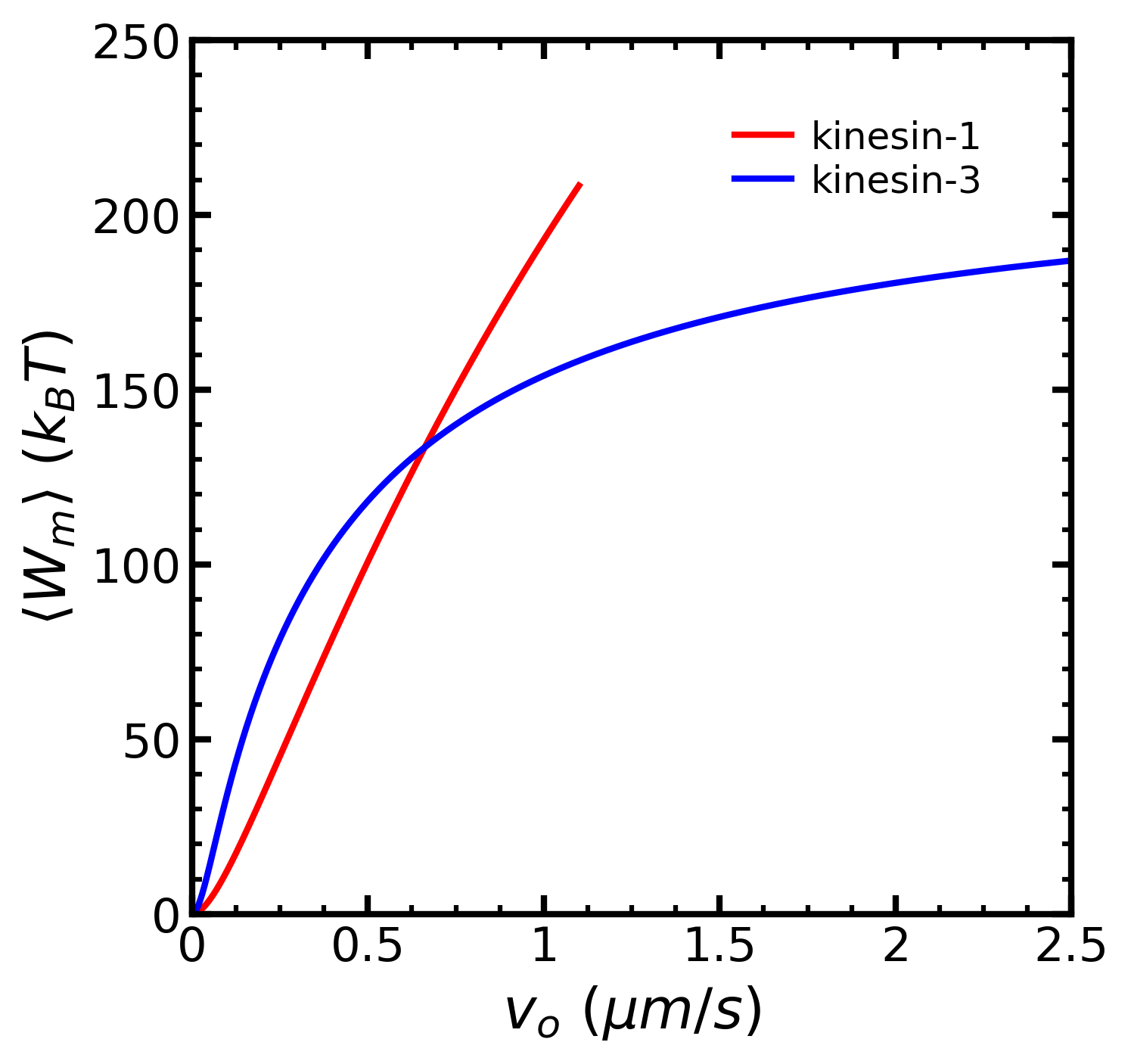}
    \caption{Work done by motor $\langle W_m\rangle$ vs $v_o$ for \textit{kinesin-1} and \textit{kinesin-3}. All parameters are same as Fig.2 of main text. }
    \label{fig:motor_work}
\end{figure}

From Eq.\ref{W_c}, we can infer, that for this protocol, $ W_c$ is necessarily negative, since the area under the curve for the case of \textit{step AB} will always be less than that of the rectangle area of side $\overline{x}^2(\tau_{1}) \tau_1$. We now use the form of $\overline{x}(t)$ (Eq.\ref{Eq:bead-av}) in the above expression to obtain Eq.6 of main text. 
Similarly, the work done by motor (Fig.\ref{fig:motor_work}) in the entire cycle can be obtained as
\begin{eqnarray}
     W_m &=& \Delta W_m^{(AB)} + \Delta W_m^{(BC)} + \Delta W_m^{(CA)}\nonumber\\
     &=& - \frac{1}{2} \frac{k_o v_{o}^{2}}{\alpha^2} \left(1 + e^{-2\alpha\tau_1} - 2 e^{-\alpha\tau_1}\right)
\end{eqnarray}

\subsection{Engine Performance in small $\alpha\langle\tau_1\rangle << 1$ and large $\alpha\langle\tau_1\rangle >> 1$ limits}
In the limit $\alpha\langle\tau_1\rangle << 1$, we Taylor expand the exponentials in Eq.6 of the main text up to cubic order in $\alpha\tau_1$. All the linear and quadratic order terms shall cancel out. Taking the average of the resulting expression over the runtime distribution (which in this limit assumes a exponential form $P(\tau_1) \rightarrow \epsilon_o e^{-\epsilon_o \tau_1}$) gives,
\begin{equation}
\langle W_c \rangle = -\frac{1}{3}\mu v_o^{2} \langle \tau_1^3 \rangle = - 2\left( \frac{\mu v_o^2}{\epsilon_o^{3}}\right) 
\label{eq:Wc-Qlow}
\end{equation}

\begin{figure}[h!]
    \centering
    \includegraphics[width =0.9\linewidth]{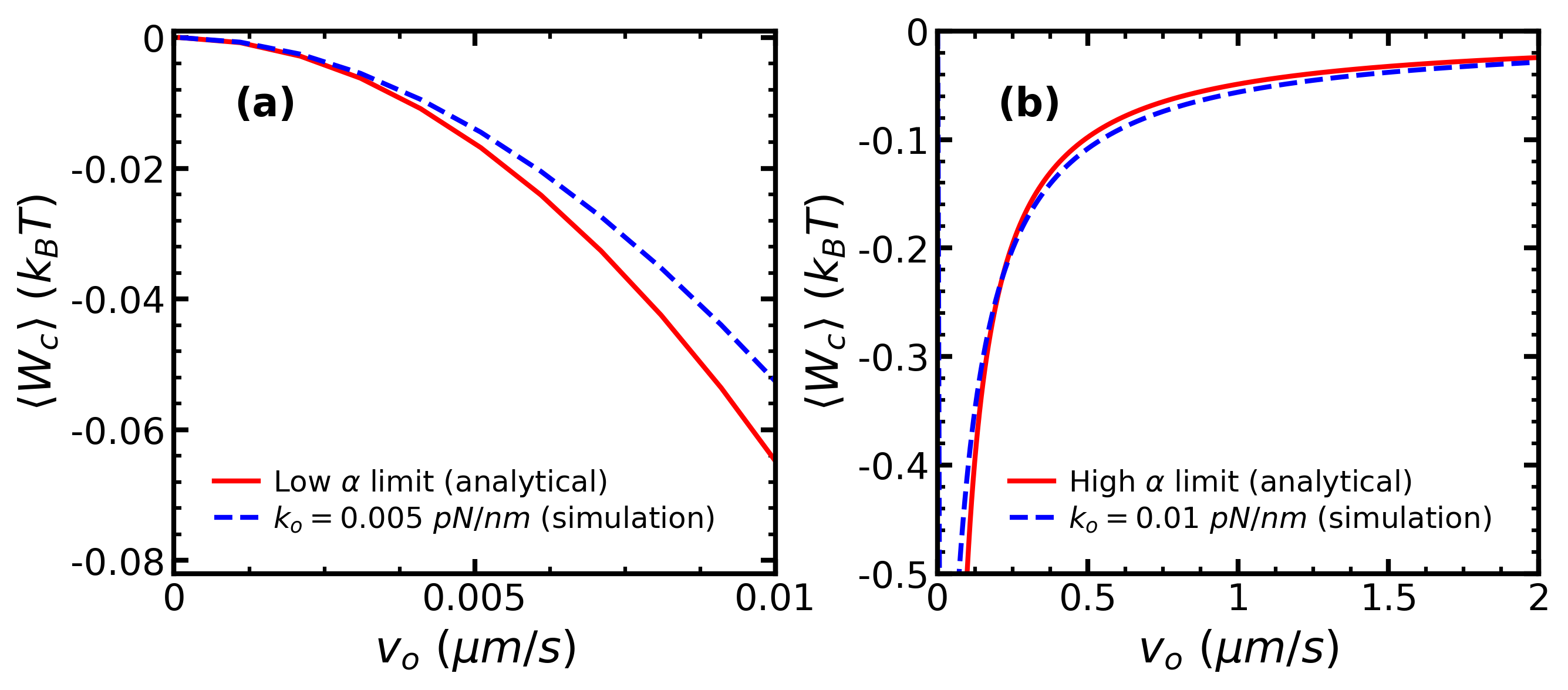}%{Fig2_new.png}
    \caption{(a) Comparison of work output $\langle W_c\rangle$ vs $v_o$ for {\it Kinesin-1} motor with Eq.\ref{eq:Wc-Qlow} corresponding to $\alpha\langle\tau_1\rangle << 1$ limit. Here, $\epsilon_o = 0.72 ~s^{-1}$, $k_o = 0.005 ~ \text{pN nm}^{-1}$ $f_s = 5.7$ pN, $f_m = 4$ pN. (b) Comparison of work output $\langle W_c \rangle$ vs $v_o$ for {\it Kinesin-3} motor with Eq.\ref{eq:Wc-Qhigh} corresponding to $\alpha\langle\tau_1\rangle >> 1$ limit. Here, $\epsilon_o = 0.23 ~s^{-1}$, $k_o = 0.1 ~ \text{pN nm}^{-1}$, $f_s = 3$ pN, $f_m = 2.7$ pN. }
    \label{fig2}
\end{figure}

In Fig.\ref{fig2}a, we compare the analytical form in Eq.\ref{eq:Wc-Qlow} with the actual value of $\langle W_c\rangle$ as $v_o$ is varied at a fixed value of $k_o$. The corresponding probability distribution function for the cumulative work $W$ is a Gaussian, with a mean value $\mu_c = - 2 N\left( \mu v_o^2 /  \epsilon_o^{3}\right)$ and variance $\sigma_c^{2} = 76 N \left(\mu^{2}v_o^4 /\epsilon_o^6 \right)$. Here $N$ is the number of independent cycles.
In the same way the expression for average work input by the motors is $\langle W_m \rangle =  \left( \frac{k_o v_o^2}{\epsilon_o^{2}}\right)$ using this, the efficiency which is defined as the ratio of work output and input, can be computed as,
\begin{equation}
\eta = \left(\frac{2 \mu }{k_o \epsilon_o} \right)
\end{equation}
The corresponding expression for average power per cycle defined as ratio of average work output to average time of the cycle is,

\begin{equation}
\langle  P_o \rangle \simeq - \frac{2\mu v_{o}^{2}\pi_o }{\epsilon_o^{2}(\epsilon_o + \pi_o)}
\end{equation}

In the same limit, it is also possible to exactly invert the work function. This allows us to obtain a closed form of work distribution. Using Eq.\ref{eq:Wc-Qlow} without taking any averages, we write $\tau_1= (-3W_c/\mu v_o^2)^{\frac{1}{3}}$. The work distribution is given by the product of Jacobian $J(\tau_1)$ and run time distribution $P(\tau_1)$ as, 

\begin{equation}
    P(W_c) =  J(\tau_1)P(\tau_1) = \frac{\epsilon_o}{\mu}\left(\frac{k_o}{f_s\alpha\tau_1}\right)^2e^{-\epsilon_o \tau_1}
\end{equation}
substitution of $\tau_1$ in terms of $W_c$ in above expression yields the Weibull distribution (Eq.9 in the main text).

\begin{equation}
P(W_c) = \frac{\epsilon_o}{3}{\left(\frac{3}{\mu v_o^2 W_c^2}\right)}^{1/3}\exp\left[ -\epsilon_o {\left(\frac{3\left|W_c\right|}{\mu v_o^2}\right)}^{1/3}\right]
\end{equation}

For the other limit where $\alpha\langle\tau_1\rangle >> 1$, by taking $\tau_1 \rightarrow \infty $ in Eq.6 of main text, it follows that
\begin{equation}
\langle W_c \rangle =  -\frac{3}{4}\frac{\mu v_o^2}{\alpha^{3}} = -\frac{3}{4}\frac{\mu f_s^3}{k_{o}^{3} v_o}
\label{eq:Wc-Qhigh}
\end{equation}
Comparison of the analytical form in Eq.\ref{eq:Wc-Qhigh} with the actual value of $\langle W_c \rangle$ as a function of $v_o$ shows good agreement for a large range of $v_o$ (See Fig.\ref{fig2}b).  

\section{Appendix E: Stochastic Simulation of single motor driven microengine in 1D}
\begin{figure}[h!]
    \centering
    \includegraphics[width=0.9\linewidth]{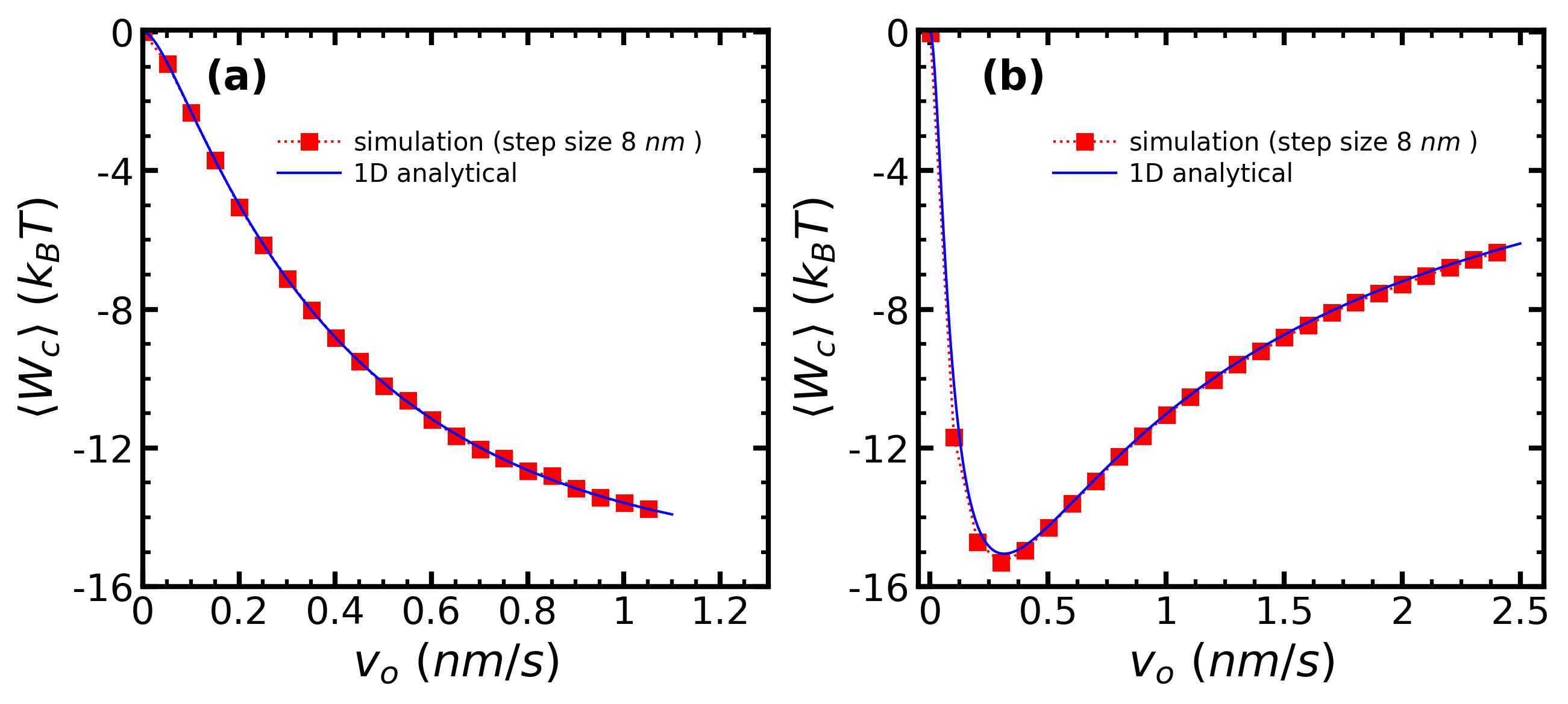}
    \caption{Comparison of average work output obtained by performing 1D Stochastic simulation with analytical expression of $\langle W_c \rangle$ in Eq.8 of the main text: $\langle W_c \rangle $ vs $v_o$ for (a) \emph{kinesin-1} and (b) \emph{kinesin-3} motor, respectively. The trap stiffness is taken as $k_o = 0.005 ~ pN nm^{-1}$. In (a) $f_s = 5.7~ pN$ , $f_m = 4.0 ~pN$ ,  $\epsilon_o = 0.72 ~ s^{-1}$, and (b) $f_s=3.0 ~ pN$ , $f_m = 2.7 ~ pN$ , $\epsilon_o = 0.23 ~ s^{-1}$. In the stochastic simulations averaging is done over $10^{6}$ independent samples.
    %(a) $\langle W_c \rangle $ vs $v_o$ For Kinesin-1 motor. Here $f_s = 5.7~ pN$ , $f_m = 4.0 ~pN$ ,  $\epsilon_o = 0.72 ~ s^{-1}$ . (b) $\langle W_c \rangle$ vs $v_o$ For kinesin-3 motor. Here $f_s=3.0 ~ pN$ , $f_m = 2.7 ~ pN$ , $\epsilon_o = 0.23 ~ s^{-1}$. For both kinesin-1 and kinesin-3 motor, $k_o = 0.005 ~ pN nm^{-1}$. In the stochastic simulations averaging is done over $10^{6}$ independent samples
    }
    \label{stochastic-work}
\end{figure}

We perform stochastic simulation in 1D for a bead that is  driven by a single kinesin motor ( with rest length $l_o = 0$).  During each simulation step of duration $\Delta t$, the motor either detaches from the MT with unbinding rate $\varepsilon(t) = \epsilon_oe^{f(t)/f_m}$ or attempts to take a step of size $d$ with a probability $P(\Delta t) = v_m \Delta t / d$, where $v_m$ is the motor velocity. The time-step is chosen as $\Delta t=10^{-4}~s$ and step size is fixed at $d=8~nm$, corresponding to step size of kinesin motor \cite{schnitzer1997nature}. The time step ensures $P(\Delta t) \ll 1$ for all values of $v_m$. The bead position is updated at every time step using the force-balance condition between the force due to optical trap and the pulling force due to motor. The force balance condition reads, $k_t \overline{x}=k_m(x_m-\overline{x})$.
%The force balance condition in this case is $k_tx_c=k_m(x_m-x_c)$. Using this relation the bead position can be determined. After every step bead position is recorded and updated.
The simulation terminates when the motor detaches from the MT, and the corresponding time is recorded as the runtime of the motor $\tau_1$. Using the values of instantaneous bead position and runtime, the work output $W_c$ is calculated for each cycle. In Fig.\ref{stochastic-work} we plot the variation of average work output $\langle W_c \rangle$ with $v_o$ obtained from stochastic simulations and compare it with the theoretical prediction of $\langle W_c \rangle$ obtained numerically. The results of stochastic simulations match excellently with the theoretical prediction. All properties are averaged over $10^6$ independent simulation runs.

\begin{figure}[h!]
    \centering
    \includegraphics[width=0.9\linewidth]{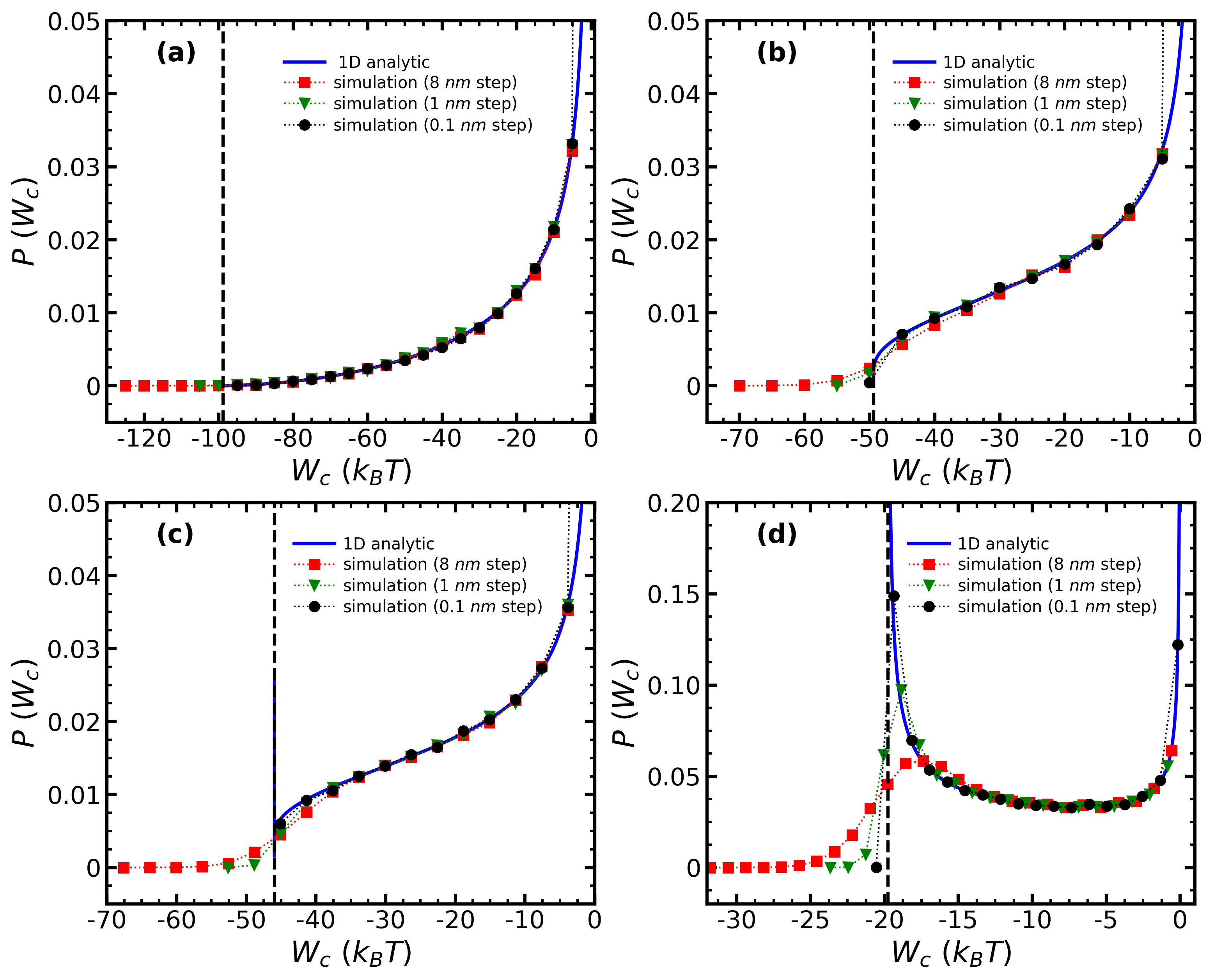}
    \caption{Comparison of $P(W_c)$ obtained from 1D stochastic simulations with theoretical prediction (Eq.7 in main text) for \emph{kinesin-3} driven microengine: (a) For $v_o=0.2 ~\mu m~ s^{-1}$, $P(W_c)$ decreases monotonically, corresponding to region I of Fig.3. (b) For $v_o = 0.4 ~\mu m~ s^{-1}$, $P(W_c)$ shows a cusp like formation, and corresponds  to region II  of Fig.3. (c) For $v_o=0.43 ~\mu m~ s^{-1}$, $P(W_c)$ exhibits a cusp and divergence at $W_c = W_{max}$. It corresponds to the region in the vicinity of the boundary separating region II with region III of Fig.3 in main text. (d) For $v_o = 1.0 ~\mu m~ s^{-1} $, the cusp is absent and the divergence of $P(W_c)$ at $W_c = W_{max}$ is observed. It corresponds to\ region III of Fig.3. Here $f_s=3.0 ~ pN$ , $f_m = 2.7 ~ pN$ , $\epsilon_o = 0.23 ~ s^{-1}$ and $k_o = 0.005 ~ pN nm^{-1}$. The corresponding value of $W_{max}$ (dashed lines) is indicated in each of the panels of the figure. Stochastic simulations are performed with $10^{6}$ independent samples.}
    \label{work-dist-1d}
\end{figure}
%\textcolor{red}{The work distributions, with step size $d=8 ~nm$ and with a continuous step (in Fig.3 of main text) are computed by utilizing $10^6$ independent cycles. It is evident from Fig.\ref{stochastic-work} that choosing a step size of $d=8~ nm$}  \textcolor{blue}{does not qualitatively affect the average work output, moreover for the continuous case, where the motor steps are treated deterministically, the corresponding distributions show an excellent match (see Fig.3 of main text).} %When the discrete motor stepping is explicitly taken into account, the distribution of work is much border, blurring out the analytic $W_{max}$ valid for deterministic case.}

When the motor variable is considered continuous, its movement can be regarded as taking place with an instantaneous step size given by $d = v_m(t)\Delta t$. Such a varying step size ensures that the probability of stepping is equal to 1 at all simulation time steps.
%Therefore, we expect that decreasing the step size, i.e. maximizing the chance of motor to take a step at each simulation instance would lead to better agreement with the analytical results.
We also compare work distributions at step sizes $d < 8~ nm$, specifically at $d=1~ nm$ and $d=0.1~ nm$. The results are presented in Fig.\ref{work-dist-1d} for kinesin-3. 
%at $k_o = 0.005 ~pN/nm$ and velocities $v_o = 0.2 ~ \mu m/s$, $0.4 ~ \mu m/s$, $0.43 ~ \mu m/s$ and $1.0~ \mu m/s$.
Indeed, it is observed that, by lowering the step size of motor (i.e. maximizing the stepping probability) results in decrease in width of distribution thereby respecting the analytic limit $W_{max}$. For smaller step size the simulation show a better match with the analytic. With increasing velocity, $P(W_c)$ shows a monotonic to a non-monotonic transition (Fig.\ref{work-dist-1d}).

%\textcolor{blue}{SD: For any velocity $v_0>1.0 ~ \mu m/s$ the time step $\Delta t$ must be choosen to be much smaller that $10^{-4}$ to ensure probability remain bounded in interval $[0,1]$ for case of $d=1$ and $d=0.1$. For $d=8$ no such requirements arise.}

%\pagebreak
\section{Appendix F: Two dimensional stochastic analysis of bead transport}
    
\subsection{Force-balance conditions}

\begin{figure}[h]
    \centering
    \includegraphics[width= 0.75\linewidth]{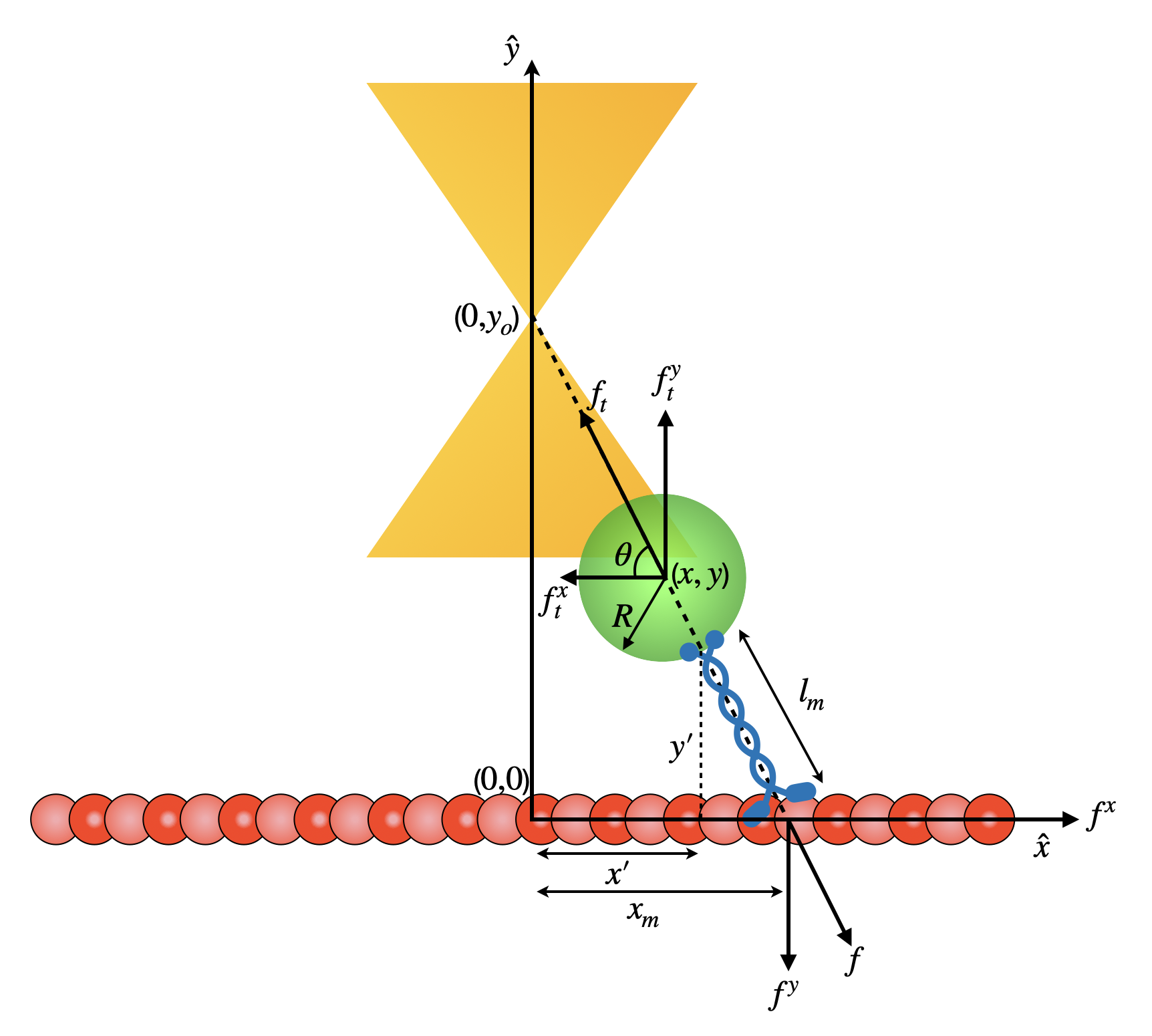}
    \caption{2D schematic of the force-balance condition for the bead transported by a single motor in optical trap. The bead has a  radius $R$ and is moving in $\hat{x}-\hat{y}$ plane. The MT is along $\hat{x}$ direction. The optical trap center is positioned at $(0,y)$ while the bead position is denoted by ($x,y$). The motor is attached to the MT at ($x_m,0$) and makes an angle $\theta$ with the MT. As the motor walks along the MT, both $x_m$ and $\theta$ changes. The other end of the motor is attached to the bead and the position of the contact between bead and the motor is denoted as $(x',y')$. The bead experiences two forces: (i) force due to motor extension ($f$) and (ii) restoring force due to optical trap ($f_t$). The components of these two forces along x and y-directions are denoted as $f^x$, $f^y$, $f_t^x$ and $f_t^y$ respectively.}
    \label{fig-schematic-appendix}
    \end{figure}

Similar to the one-dimensional analysis, the motor in this case is modeled as a harmonic spring with a spring constant $k_m$. As the motor progresses along the underlying MT filament, it exerts a pulling force on the bead. This force is counteracted by the restoring force arising from the optical trap potential. Both the forces acting on the bead and its displacement vector from the center of the optical trap typically have components in two directions: the horizontal direction, aligned with the MT axis, and the vertical direction, perpendicular to it (Fig. \ref{fig-schematic-appendix}).\\

%    We model the motor as a harmonic spring with spring constant , $k_m$. For this system, when the motor is attached to the underlying MT filament, the bead experience a pulling force due the motor and an opposing restoring force due the harmonic potential of the optical trap. In general, the displacement vector of the bead  from the optical trap center has components in the horizontal direction (along the axis of MT ) as well along the vertical direction (see Fig.1a). 
 
 %While many previous studies on the motor transport in optical trap setting have only considered one-dimensional transport of the bead \cite{kunwar2011pnas, rai2013cell, klumpp2005pnas}, ignoring the vertical component of the forces due to the motor and the optical trap, in reality, the bead is subjected to force along the vertical direction as well \cite{svoboda1994cell,pyrpassopoulos2020biophysj,khataee2019prl} (see Fig. \ref{fig:schematic}b).
%The effect of the optical trap on the cellular bead is modelled as a harmonic potential. While the ratio of trap stiffness in the horizontal and vertical directions varies in range of $1.5-4$ depending on the choice of the bead size and its coating material \cite{ashkin1992biophysj,bormuth2008opticsexp}, 
%for simplicity, we assume that the harmonic potential is isotropic with a spring constant $k_t$. 

We consider that the microtubule (MT) lies along the X-axis at $y=0$, with the optical trap center located at 
$(0,y_o)$. Here, $y_o=l_o+R$, where $l_o$ represents the rest length of the motor, and 
$R$ is the radius of the bead. At $t=0$, the motor is attaches to the MT, and the bead is positioned at the trap center. Consequently, we assume that at $t=0$, the motor is vertically aligned at $x_m=0$, and its length equals its rest length ($l_o$).

%For simplicity, we consider the MT to be along X-axis at $y_{MT}=0$ and the optical trap center positioned at $(0,y_o)$. We choose $y_o=l_o+R$, where $l_o$ is the restlength of the motor and $R$ is the radius of the bead (bead). We assume that at $t=0$, the motor is attached to the MT and the bead is at the trap center. Therefore, at $t=0$, the motor is vertically positioned at $x_m=0$ and the length of the motor is equal to its restlength ($l_o$). \\

%We consider a bead  of radius $R$. The coordinate system is chosen such that the optical trap center is at position $(0,y_o)$ while MT is aligned along $y=0$.  The initial bead position is at the trap center. We assume that the motor attaches to the MT at (0,0) at time $t=0$. Molecular motors  have a finite rest length $l_o$. For kinesin motor the $l_o \simeq 110 nm$. Intially at $t=0$, the motor position is $y_o=R+l_o$. %such that the motor is in a vertical position with an extension equal to $l_o$. 

As the motor moves along the microtubule (MT), it stretches beyond its rest length, generating a restoring force that acts on the bead in the direction of the motor's extension. This restoring force causes the bead to deviate from the trap center, thereby activating the optical trap force, which pulls the bead back toward the trap center. The equilibrium position of the bead and the orientation of the motor - characterized by the angle 
$\theta$ formed between the motor head and the MT - are determined by the balance between these opposing forces and the torques they produce. The resulting torques cause the bead to rotate about its center, reaching equilibrium when the motor aligns along the line connecting the bead center to the trap center. Under these conditions, the force balance can be expressed as

\begin{eqnarray}
  k_t^x x &=& k_m\Theta(l_m-l_o) \left[ (x_m-x') - l_o\frac{(x'-x)}{R}\right] \label{eq:xc} \\
    k_t^y(y_o&-&y) = k_m\Theta(l_m-l_o) \left[y'-l_o\frac{(y-y')}{R}\right] \label{eq:yc}
\end{eqnarray}
where $x_m$ is the motor position on the MT, $l_m$ is the length of the motor, $(x', y')$ is the motor binding position on the bead surface and $(x,y)$ is the position of the bead center, $\Theta$ is the Heavy-side theta function, $k_t^x$ and $k_t^y$ are the trap stiffness along horizontal and vertical directions, respectively (Fig.\ref{fig-schematic-appendix}) \cite{sundararajan2024theoretical}. Note that the Heavy-side $\Theta$ function ensures that the motor exerts a force on the bead only when it is stretched beyond its rest length (i.e. $l_m>l_o$).\\

It is important to note that the force balance condition between the motor force and the trap force in the vertical direction (Eq.\ref{eq:yc}) remains valid as long as the bead does not come into contact with the underlying MT. Once the motor moves along the MT and the bead is in contact with the MT, an additional normal force (generated by the MT in the vertical direction) will act on the bead \cite{fisher2005pnas,khataee2019prl,pyrpassopoulos2020biophysj}. In such a scenario, the vertical force balance will be determined by this normal force in addition to the trap force and the motor force in the vertical direction. Consequently, when the bead touches the MT, only the horizontal force balance condition (Eq. \ref{eq:xc}) should be considered.

From a geometric perspective, the motor binding position $(x',y')$ must satisfy the following equation -

\begin{equation}
    (x'-x)^2+(y'-y)^2 = R^2 \label{eq:surface}
\end{equation}

Therefore, at equilibrium,
\begin{equation}
    tan\theta = \frac{y'}{x_m-x'} = \frac{y}{x_m-x} \label{eq:triangle}
\end{equation}

By simultaneously solving these four equations (Eqs. \ref{eq:xc}-\ref{eq:triangle}), the values of $x,y,x',y'$ can be determined for a given motor position $x_m$.

\subsection{Stochastic simulation of single motor driven microengine in 2D}

We use force-balance conditions and geometric constraints (Eqs.\ref{eq:xc}-\ref{eq:triangle}) to perform stochastic simulations of single kinesin-driven bead transport in an optical trap. The force $(f)$ acting on the motor at any moment is given by $f = k_m(l_m - l_o)$, where $l_m > l_o$ and $l_m = \sqrt{(x_m - x')^2 + y'^2}$. The motor's velocity under load follows the relationship $v_m = v_o(1 - f/f_s)$, where $v_o$ is the motor's velocity without load, and $f_s$ is its stall force. Similar to 1D simulations, at each time step, the motor either detaches from the MT with an unbinding rate $\epsilon$ or attempts to take a step forward on the MT. Here also we chose  $\Delta t=10^{-4} ~s$ and $d=8 ~nm$ corresponding to kinesin step length \cite{schnitzer1997nature}. After each step, the force-balance conditions are recalculated to update the values of $x$, $y$, $x'$, and $y'$ based on Eqs. \ref{eq:xc}-\ref{eq:triangle}. The simulation terminates when the motor detaches from the MT.\\

The expressions of work outputs for paths AB and BC can be written as follows
\begin{eqnarray}
    \Delta W^{(AB)}_{c} &=& \int_{o}^{\tau_1}\left( \frac{\partial U}{\partial \vec{k_t}}\right) \cdot \dot {\vec{k_t}} dt \nonumber\\
    &=& \frac{1}{2} \int_{o}^{\tau_1} \left[\mu_x\overline x^{2} + \mu_y\left(y_o-\overline{y}\right)^2\right]dt\\
    \Delta W^{(BC)}_c &=& -\frac{1}{2}\tau_1 \left[\mu_x\overline{x}^2(\tau_1)+\mu_y\left\{y_o-\overline{y}(\tau_1)\right\}^2\right]
\end{eqnarray}
where $\vec{k}_t = k_t^x\hat{x} + k_t^y\hat{y}=(k_o^x+\mu_xt)\hat{x}+(k_o^y+\mu_yt)\hat{y}$ is a two-dimensional vector and
$\mu_x$ and $\mu_y$ denote the rates of change of the trap
stiffness along the MT direction and perpendicular to the MT direction, respectively.  In this study, we have always taken $k_o^y=k_o^x/3$ \cite{ashkin1992biophysj,bormuth2008opticsexp}, $\mu_x = k_o^x/10$ and $\mu_y = k_o^y/10$. All properties are averaged over $10^6$ independent simulation runs. \\

%All properties are averaged over $10^6$ independent simulation runs. Fig.S2 and Fig.S3 displays the results obtained from stochastic simulations and compares it with the theoretical analytical results predicted by considering the bead-motor complex in optical trap as an effective 1D system.

\section{Appendix G: Effect of time delay in feedback process} If there is a time delay of $\delta t_a$ in the feedback process from the instant of motor attachment at $t=0$, then the stiffness of the optical trap continues to remain $k_o$ for a duration of $\delta t_a$ even after the motor has attached to the MT. Therefore, the change in the value of the trap stiffness until the motor detaches is $\Delta k = \mu (\tau_1 - \delta t_a)$. Using the expression for work output per cycle $W_c$  (Eq.\ref{eq:Wc-Qlow}),  in the limit of $\alpha \langle \tau_1 \rangle << 1$,  we can estimate the reduction of the total work output per cycle, $
\delta W_{c}^a \simeq \delta t_a \left( \frac{\mu v_o^{2}}{2 \epsilon_o^2} \right)$.

In order to estimate the reduction of work output due to the delay in feedback process at the motor detachment step of the engine cycle,  we proceed as follows: If the motor is already detached, the position of the bead relaxes from the original position $x(\tau_1)$ to a value $ x(\tau_1)e^{-\frac{k_o}{\gamma} \delta t_d} \simeq x(\tau_1)( 1 - \frac{k_o}{\gamma} \delta t_d)$. Again, using the expression for $\langle W_c\rangle $, (in the limit of $\alpha\langle \tau_1 \rangle << 1$),  we can estimate the reduction of work output per cycle as, 
$\delta W_{c}^{d} \simeq \delta t_d \left[ \mu v_o^{2} \langle \tau_1^{3}\rangle \left(\frac{k_o}{\gamma}\right)\right]$.
Beyond a critical time delay $\delta t_c$, the net work output would be zero and the engine would cease to function. Then it follows that, $
\delta t_c = \frac{1}{3} \left( \frac{\gamma}{k_o}\right)$. Thus, for a delay beyond typical relaxation time for the bead in the optical trap ($\gamma/k_o$), no useful work can be extracted from the engine and it sets a bound for the performance of the engine.
The delay time in the feedback protocol, $\delta t_f$ has to be such that, $\delta t_f << \delta t_d$, for the engine performance to remain robust.  %For a bead of micron size diameter, for a trap stiffness of $k_o = 10^{-2}$ $ pN nm^{-1}$, the bead relaxation time scale, $ \gamma/k_o \sim 10^{-3}$ s. 

\bibliographystyle{unsrtnat}
\bibliography{paper-engine-v7Notes}

\begin{thebibliography}{58}
\providecommand{\natexlab}[1]{#1}
\providecommand{\url}[1]{\texttt{#1}}
\expandafter\ifx\csname urlstyle\endcsname\relax
  \providecommand{\doi}[1]{doi: #1}\else
  \providecommand{\doi}{doi: \begingroup \urlstyle{rm}\Url}\fi

\bibitem[Howard and Clark(2002)]{howard2002mechanics}
Jonathon Howard and RL~Clark.
\newblock Mechanics of motor proteins and the cytoskeleton.
\newblock \emph{Appl. Mech. Rev.}, 55\penalty0 (2):\penalty0 B39--B39, 2002.

\bibitem[Phillips et~al.(2012)Phillips, Kondev, Theriot, and
  Garcia]{phillips2012physical}
Rob Phillips, Jane Kondev, Julie Theriot, and Hernan Garcia.
\newblock \emph{Physical biology of the cell}.
\newblock Garland Science, 2012.

\bibitem[Oster(2002)]{oster2002brownian}
George Oster.
\newblock Brownian ratchets: Darwin's motors.
\newblock \emph{Nature}, 417\penalty0 (6884):\penalty0 25--25, 2002.

\bibitem[J{\"u}licher et~al.(1997)J{\"u}licher, Ajdari, and
  Prost]{julicher1997modeling}
Frank J{\"u}licher, Armand Ajdari, and Jacques Prost.
\newblock Modeling molecular motors.
\newblock \emph{Reviews of Modern Physics}, 69\penalty0 (4):\penalty0 1269,
  1997.

\bibitem[Reimann(2002)]{reimann2002brownian}
Peter Reimann.
\newblock Brownian motors: noisy transport far from equilibrium.
\newblock \emph{Physics reports}, 361\penalty0 (2-4):\penalty0 57--265, 2002.

\bibitem[Ait-Haddou and Herzog(2003)]{ait2003brownian}
Rachid Ait-Haddou and Walter Herzog.
\newblock Brownian ratchet models of molecular motors.
\newblock \emph{Cell biochemistry and biophysics}, 38\penalty0 (2):\penalty0
  191--213, 2003.

\bibitem[Blickle and Bechinger(2012)]{blickle2012realization}
Valentin Blickle and Clemens Bechinger.
\newblock Realization of a micrometre-sized stochastic heat engine.
\newblock \emph{Nature Physics}, 8\penalty0 (2):\penalty0 143--146, 2012.

\bibitem[Mart{\'\i}nez et~al.(2016)Mart{\'\i}nez, Rold{\'a}n, Dinis, Petrov,
  Parrondo, and Rica]{martinez2016brownian}
Ignacio~A Mart{\'\i}nez, {\'E}dgar Rold{\'a}n, Luis Dinis, Dmitri Petrov,
  Juan~MR Parrondo, and Ra{\'u}l~A Rica.
\newblock Brownian carnot engine.
\newblock \emph{Nature physics}, 12\penalty0 (1):\penalty0 67--70, 2016.

\bibitem[Ciliberto(2017)]{ciliberto2017experiments}
Sergio Ciliberto.
\newblock Experiments in stochastic thermodynamics: Short history and
  perspectives.
\newblock \emph{Physical Review X}, 7\penalty0 (2):\penalty0 021051, 2017.

\bibitem[Krishnamurthy et~al.(2016)Krishnamurthy, Ghosh, Chatterji, Ganapathy,
  and Sood]{krishnamurthy2016micrometre}
Sudeesh Krishnamurthy, Subho Ghosh, Dipankar Chatterji, Rajesh Ganapathy, and
  AK~Sood.
\newblock A micrometre-sized heat engine operating between bacterial
  reservoirs.
\newblock \emph{Nature Physics}, 12\penalty0 (12):\penalty0 1134--1138, 2016.

\bibitem[Krishnamurthy et~al.(2023)Krishnamurthy, Ganapathy, and
  Sood]{krishnamurthy2023overcoming}
Sudeesh Krishnamurthy, Rajesh Ganapathy, and AK~Sood.
\newblock Overcoming power-efficiency tradeoff in a micro heat engine by
  engineered system-bath interactions.
\newblock \emph{Nature Communications}, 14\penalty0 (1):\penalty0 6842, 2023.

\bibitem[Roy et~al.(2021)Roy, Leroux, Sood, and Ganapathy]{roy2021tuning}
Niloyendu Roy, Nathan Leroux, AK~Sood, and Rajesh Ganapathy.
\newblock Tuning the performance of a micrometer-sized stirling engine through
  reservoir engineering.
\newblock \emph{Nature Communications}, 12\penalty0 (1):\penalty0 4927, 2021.

\bibitem[Saha and Marathe(2019)]{saha2019stochastic}
Arnab Saha and Rahul Marathe.
\newblock Stochastic work extraction in a colloidal heat engine in the presence
  of colored noise.
\newblock \emph{Journal of Statistical Mechanics: Theory and Experiment},
  2019\penalty0 (9):\penalty0 094012, 2019.

\bibitem[Szilard(1929)]{szilard1929entropieverminderung}
Leo Szilard.
\newblock {\"U}ber die entropieverminderung in einem thermodynamischen system
  bei eingriffen intelligenter wesen.
\newblock \emph{Zeitschrift f{\"u}r Physik}, 53\penalty0 (11):\penalty0
  840--856, 1929.

\bibitem[Parrondo et~al.(2015)Parrondo, Horowitz, and
  Sagawa]{parrondo2015thermodynamics}
Juan~MR Parrondo, Jordan~M Horowitz, and Takahiro Sagawa.
\newblock Thermodynamics of information.
\newblock \emph{Nature physics}, 11\penalty0 (2):\penalty0 131--139, 2015.

\bibitem[Cao and Feito(2009)]{cao2009thermodynamics}
Francisco~J Cao and M~Feito.
\newblock Thermodynamics of feedback controlled systems.
\newblock \emph{Physical Review E-Statistical, Nonlinear, and Soft Matter
  Physics}, 79\penalty0 (4):\penalty0 041118, 2009.

\bibitem[Tohme et~al.(2024)Tohme, Bedoya, di~Bello, Bresque, Manzano, and
  Rold{\'a}n]{tohme2024gambling}
Tarek Tohme, Valentina Bedoya, Costantino di~Bello, L{\'e}a Bresque, Gonzalo
  Manzano, and {\'E}dgar Rold{\'a}n.
\newblock Gambling carnot engine.
\newblock \emph{arXiv preprint arXiv:2409.17212}, 2024.

\bibitem[du~Buisson et~al.(2024)du~Buisson, Sivak, and
  Bechhoefer]{du2024performance}
Johan du~Buisson, David~A Sivak, and John Bechhoefer.
\newblock Performance limits of information engines.
\newblock \emph{Advances in Physics: X}, 9\penalty0 (1):\penalty0 2352112,
  2024.

\bibitem[Saha et~al.(2023)Saha, Ehrich, Gavrilov, Still, Sivak, and
  Bechhoefer]{saha2023information}
Tushar~K Saha, Jannik Ehrich, Mom{\v{c}}ilo Gavrilov, Susanne Still, David~A
  Sivak, and John Bechhoefer.
\newblock Information engine in a nonequilibrium bath.
\newblock \emph{Physical Review Letters}, 131\penalty0 (5):\penalty0 057101,
  2023.

\bibitem[Malgaretti and Stark(2022)]{malgaretti2022szilard}
Paolo Malgaretti and Holger Stark.
\newblock Szilard engines and information-based work extraction for active
  systems.
\newblock \emph{Physical review letters}, 129\penalty0 (22):\penalty0 228005,
  2022.

\bibitem[Paneru et~al.(2018)Paneru, Lee, Tlusty, and Pak]{paneru2018lossless}
Govind Paneru, Dong~Yun Lee, Tsvi Tlusty, and Hyuk~Kyu Pak.
\newblock Lossless brownian information engine.
\newblock \emph{Physical review letters}, 120\penalty0 (2):\penalty0 020601,
  2018.

\bibitem[Ribezzi-Crivellari and Ritort(2019)]{ribezzi2019large}
Marco Ribezzi-Crivellari and Felix Ritort.
\newblock Large work extraction and the landauer limit in a continuous maxwell
  demon.
\newblock \emph{Nature Physics}, 15\penalty0 (7):\penalty0 660--664, 2019.

\bibitem[Paneru et~al.(2022)Paneru, Dutta, and Pak]{paneru2022colossal}
Govind Paneru, Sandipan Dutta, and Hyuk~Kyu Pak.
\newblock Colossal power extraction from active cyclic brownian information
  engines.
\newblock \emph{The Journal of Physical Chemistry Letters}, 13\penalty0
  (30):\penalty0 6912--6918, 2022.

\bibitem[Leff and Rex(2002)]{leff2002maxwell}
Harvey Leff and Andrew~F Rex.
\newblock \emph{Maxwell's Demon 2 Entropy, Classical and Quantum Information,
  Computing}.
\newblock CRC Press, 2002.

\bibitem[Schmiedl and Seifert(2007)]{schmiedl2007efficiency}
Tim Schmiedl and Udo Seifert.
\newblock Efficiency at maximum power: An analytically solvable model for
  stochastic heat engines.
\newblock \emph{Europhysics letters}, 81\penalty0 (2):\penalty0 20003, 2007.

\bibitem[Rana et~al.(2014)Rana, Pal, Saha, and Jayannavar]{rana2014single}
Shubhashis Rana, PS~Pal, Arnab Saha, and AM~Jayannavar.
\newblock Single-particle stochastic heat engine.
\newblock \emph{Physical review E}, 90\penalty0 (4):\penalty0 042146, 2014.

\bibitem[Verley et~al.(2014{\natexlab{a}})Verley, Willaert, Van~den Broeck, and
  Esposito]{verley2014universal}
Gatien Verley, Tim Willaert, Christian Van~den Broeck, and Massimiliano
  Esposito.
\newblock Universal theory of efficiency fluctuations.
\newblock \emph{Physical Review E}, 90\penalty0 (5):\penalty0 052145,
  2014{\natexlab{a}}.

\bibitem[Verley et~al.(2014{\natexlab{b}})Verley, Esposito, Willaert, and
  Van~den Broeck]{verley2014unlikely}
Gatien Verley, Massimiliano Esposito, Tim Willaert, and Christian Van~den
  Broeck.
\newblock The unlikely carnot efficiency.
\newblock \emph{Nature communications}, 5\penalty0 (1):\penalty0 4721,
  2014{\natexlab{b}}.

\bibitem[Roy et~al.(2023)Roy, Sood, and Ganapathy]{prlroy}
Niloyendu Roy, A.~K. Sood, and Rajesh Ganapathy.
\newblock Harnessing viscoelasticity to suppress irreversibility buildup in a
  colloidal stirling engine.
\newblock \emph{Phys. Rev. Lett.}, 131:\penalty0 238201, 2023.

\bibitem[Sundararajan et~al.(2024)Sundararajan, Guha, Muhuri, and
  Mitra]{sundararajan2024theoretical}
Naren Sundararajan, Sougata Guha, Sudipto Muhuri, and Mithun~K Mitra.
\newblock Theoretical analysis of cargo transport by catch bonded motors in
  optical trapping assays.
\newblock \emph{Soft Matter}, 20\penalty0 (3):\penalty0 566--577, 2024.

\bibitem[Kunwar et~al.(2011)Kunwar, Tripathy, Xu, Mattson, Anand, Sigua,
  Vershinin, McKenney, Yu, Mogilner, et~al.]{kunwar2011mechanical}
Ambarish Kunwar, Suvranta~K Tripathy, Jing Xu, Michelle~K Mattson, Preetha
  Anand, Roby Sigua, Michael Vershinin, Richard~J McKenney, Clare~C Yu,
  Alexander Mogilner, et~al.
\newblock Mechanical stochastic tug-of-war models cannot explain bidirectional
  lipid-droplet transport.
\newblock \emph{Proceedings of the National Academy of Sciences}, 108\penalty0
  (47):\penalty0 18960--18965, 2011.

\bibitem[Ashkin(1992)]{ashkin1992biophysj}
Arthur Ashkin.
\newblock Forces of a single-beam gradient laser trap on a dielectric sphere in
  the ray optics regime.
\newblock \emph{Biophysical journal}, 61\penalty0 (2):\penalty0 569--582, 1992.

\bibitem[Fisher and Kim(2005)]{fisher2005pnas}
Michael~E Fisher and Young~C Kim.
\newblock Kinesin crouches to sprint but resists pushing.
\newblock \emph{Proceedings of the National Academy of Sciences}, 102\penalty0
  (45):\penalty0 16209--16214, 2005.

\bibitem[Khataee and Howard(2019)]{khataee2019prl}
Hamid Khataee and Jonathon Howard.
\newblock Force generated by two kinesin motors depends on the load direction
  and intermolecular coupling.
\newblock \emph{Physical review letters}, 122\penalty0 (18):\penalty0 188101,
  2019.

\bibitem[Pyrpassopoulos et~al.(2020)Pyrpassopoulos, Shuman, and
  Ostap]{pyrpassopoulos2020biophysj}
Serapion Pyrpassopoulos, Henry Shuman, and E~Michael Ostap.
\newblock Modulation of kinesin’s load-bearing capacity by force geometry and
  the microtubule track.
\newblock \emph{Biophysical journal}, 118\penalty0 (1):\penalty0 243--253,
  2020.

\bibitem[Coppin et~al.(1997)Coppin, Pierce, Hsu, and Vale]{coppin1997load}
Chris~M Coppin, Daniel~W Pierce, Long Hsu, and Ronald~D Vale.
\newblock The load dependence of kinesin’s mechanical cycle.
\newblock \emph{Proceedings of the National Academy of Sciences}, 94\penalty0
  (16):\penalty0 8539--8544, 1997.

\bibitem[Klumpp and Lipowsky(2005)]{klumpp2005cooperative}
Stefan Klumpp and Reinhard Lipowsky.
\newblock Cooperative cargo transport by several molecular motors.
\newblock \emph{Proceedings of the National Academy of Sciences}, 102\penalty0
  (48):\penalty0 17284--17289, 2005.

\bibitem[M{\"u}ller et~al.(2008)M{\"u}ller, Klumpp, and
  Lipowsky]{muller2008tug}
Melanie~JI M{\"u}ller, Stefan Klumpp, and Reinhard Lipowsky.
\newblock Tug-of-war as a cooperative mechanism for bidirectional cargo
  transport by molecular motors.
\newblock \emph{Proceedings of the National Academy of Sciences}, 105\penalty0
  (12):\penalty0 4609--4614, 2008.

\bibitem[Puri et~al.(2019)Puri, Gupta, Chandel, Naskar, Nair, Chaudhuri, Mitra,
  and Muhuri]{puri2019dynein}
Palka Puri, Nisha Gupta, Sameep Chandel, Supriyo Naskar, Anil Nair, Abhishek
  Chaudhuri, Mithun~K Mitra, and Sudipto Muhuri.
\newblock Dynein catch bond as a mediator of codependent bidirectional cellular
  transport.
\newblock \emph{Physical Review Research}, 1\penalty0 (2):\penalty0 023019,
  2019.

\bibitem[Svoboda and Block(1994)]{svoboda1994force}
Karel Svoboda and Steven~M Block.
\newblock Force and velocity measured for single kinesin molecules.
\newblock \emph{Cell}, 77\penalty0 (5):\penalty0 773--784, 1994.

\bibitem[Visscher et~al.(1999)Visscher, Schnitzer, and Block]{Visscher1999}
Koen Visscher, Mark~J. Schnitzer, and Steven~M. Block.
\newblock Single kinesin molecules studied with a molecular force clamp.
\newblock \emph{Nature}, 400\penalty0 (6740):\penalty0 184--189, 1999.

\bibitem[tim()]{timescale}
For sub-micron size bead, $\tau_c \sim 10^{-18}~s$, $\tau_b \sim 10^{-4}~ s$ (
  for $k_o = 0.005~ pN nm^{-1}$) and $\tau_m$ has a typical range of $\tau_m
  \sim (10^{-2} - 10^{1})~ s $.

\bibitem[Brenner et~al.(2020)Brenner, Berger, Rao, Nicholas, and
  Gennerich]{brenner2020force}
Sibylle Brenner, Florian Berger, Lu~Rao, Matthew~P Nicholas, and Arne
  Gennerich.
\newblock Force production of human cytoplasmic dynein is limited by its
  processivity.
\newblock \emph{Science advances}, 6\penalty0 (15):\penalty0 eaaz4295, 2020.

\bibitem[Leduc et~al.(2004)Leduc, Camp{\`a}s, Zeldovich, Roux, Jolimaitre,
  Bourel-Bonnet, Goud, Joanny, Bassereau, and Prost]{leduc2004cooperative}
C{\'e}cile Leduc, Otger Camp{\`a}s, Konstantin~B Zeldovich, Aur{\'e}lien Roux,
  Pascale Jolimaitre, Line Bourel-Bonnet, Bruno Goud, Jean-Fran{\c{c}}ois
  Joanny, Patricia Bassereau, and Jacques Prost.
\newblock Cooperative extraction of membrane nanotubes by molecular motors.
\newblock \emph{Proceedings of the National Academy of Sciences}, 101\penalty0
  (49):\penalty0 17096--17101, 2004.

\bibitem[Peliti and Pigolotti(2021)]{peliti2021stochastic}
Luca Peliti and Simone Pigolotti.
\newblock \emph{Stochastic thermodynamics: an introduction}.
\newblock Princeton University Press, 2021.

\bibitem[Soppina et~al.(2022)Soppina, Patel, Shewale, Rai, Sivaramakrishnan,
  Naik, and Soppina]{soppina2022kinesin}
Pushpanjali Soppina, Nishaben Patel, Dipeshwari~J Shewale, Ashim Rai, Sivaraj
  Sivaramakrishnan, Pradeep~K Naik, and Virupakshi Soppina.
\newblock Kinesin-3 motors are fine-tuned at the molecular level to endow
  distinct mechanical outputs.
\newblock \emph{BMC biology}, 20\penalty0 (1):\penalty0 177, 2022.

\bibitem[Guo et~al.(2019)Guo, Shi, Wang, and Xie]{guo2019force}
Si-Kao Guo, Xiao-Xuan Shi, Peng-Ye Wang, and Ping Xie.
\newblock Force dependence of unbinding rate of kinesin motor during its
  processive movement on microtubule.
\newblock \emph{Biophysical chemistry}, 253:\penalty0 106216, 2019.

\bibitem[Budaitis et~al.(2021)Budaitis, Jariwala, Rao, Yue, Sept, Verhey, and
  Gennerich]{budaitis2021pathogenic}
Breane~G Budaitis, Shashank Jariwala, Lu~Rao, Yang Yue, David Sept, Kristen~J
  Verhey, and Arne Gennerich.
\newblock Pathogenic mutations in the kinesin-3 motor kif1a diminish force
  generation and movement through allosteric mechanisms.
\newblock \emph{Journal of Cell Biology}, 220\penalty0 (4):\penalty0
  e202004227, 2021.

\bibitem[Rai et~al.(2013)Rai, Rai, Ramaiya, Jha, and Mallik]{rai2013molecular}
Arpan~K Rai, Ashim Rai, Avin~J Ramaiya, Rupam Jha, and Roop Mallik.
\newblock Molecular adaptations allow dynein to generate large collective
  forces inside cells.
\newblock \emph{Cell}, 152\penalty0 (1):\penalty0 172--182, 2013.

\bibitem[Rosing and Slater(1972)]{rosing1972value}
J~Rosing and EC~Slater.
\newblock The value of $\delta$g° for the hydrolysis of atp.
\newblock \emph{Biochimica et Biophysica Acta (BBA)-Bioenergetics},
  267\penalty0 (2):\penalty0 275--290, 1972.

\bibitem[Schnitzer and Block(1997)]{schnitzer1997nature}
Mark~J Schnitzer and Steven~M Block.
\newblock Kinesin hydrolyses one atp per 8-nm step.
\newblock \emph{Nature}, 388\penalty0 (6640):\penalty0 386--390, 1997.

\bibitem[Howard et~al.(1989)Howard, Hudspeth, and Vale]{howard1989movement}
J~Howard, AJ~Hudspeth, and RD~Vale.
\newblock Movement of microtubules by single kinesin molecules.
\newblock \emph{Nature}, 342\penalty0 (6246):\penalty0 154--158, 1989.

\bibitem[Erickson et~al.(2011)Erickson, Jia, Gross, and
  Yu]{erickson2011molecular}
Robert~P Erickson, Zhiyuan Jia, Steven~P Gross, and Clare~C Yu.
\newblock How molecular motors are arranged on a cargo is important for
  vesicular transport.
\newblock \emph{PLoS computational biology}, 7\penalty0 (5):\penalty0 e1002032,
  2011.

\bibitem[U{\c{c}}ar and Lipowsky(2019)]{uccar2019force}
Mehmet~Can U{\c{c}}ar and Reinhard Lipowsky.
\newblock Force sharing and force generation by two teams of elastically
  coupled molecular motors.
\newblock \emph{Scientific reports}, 9\penalty0 (1):\penalty0 454, 2019.

\bibitem[Bormuth et~al.(2008)Bormuth, Jannasch, Ander, van Kats, van Blaaderen,
  Howard, and Sch{\"a}ffer]{bormuth2008opticsexp}
Volker Bormuth, Anita Jannasch, Marcel Ander, Carlos~M van Kats, Alfons van
  Blaaderen, Jonathon Howard, and Erik Sch{\"a}ffer.
\newblock Optical trapping of coated microspheres.
\newblock \emph{Optics express}, 16\penalty0 (18):\penalty0 13831--13844, 2008.

\bibitem[Shannon and Weaver(1998)]{shannon1998mathematical}
Claude~E Shannon and Warren Weaver.
\newblock \emph{The mathematical theory of communication}.
\newblock University of Illinois press, 1998.

\bibitem[Thomas and Joy(2006)]{thomas2006elements}
MTCAJ Thomas and A~Thomas Joy.
\newblock \emph{Elements of information theory}.
\newblock Wiley-Interscience, 2006.

\bibitem[Guha et~al.(2021)Guha, Mitra, Pagonabarraga, and
  Muhuri]{guha2021novel}
Sougata Guha, Mithun~K Mitra, Ignacio Pagonabarraga, and Sudipto Muhuri.
\newblock Novel mechanism for oscillations in catchbonded motor-filament
  complexes.
\newblock \emph{Biophysical Journal}, 120\penalty0 (18):\penalty0 4129--4136,
  2021.

\end{thebibliography}

\end{document}